\documentclass[12pt]{article}
\usepackage{etoolbox}
\usepackage[utf8]{inputenc}
\usepackage[top = 2 cm, bottom = 2 cm, left = 2.5 cm, right = 1.5 cm]{geometry}
\usepackage[]{amsmath}
\usepackage[]{amssymb}
\usepackage[pdftex]{graphicx}
\graphicspath{ {images/} }
\usepackage{float}
\restylefloat{figure} 
\usepackage{expl3}
\usepackage{setspace}
\usepackage{subfig}
\usepackage{indentfirst}
\usepackage{amsthm}
\usepackage{amsfonts}
\usepackage{mathrsfs}
\usepackage{mathtools}
\usepackage{cmap}
\usepackage{exercise}
\usepackage{tikz}
\usepackage{ytableau}
\usetikzlibrary{tikzmark,calc}
\usetikzlibrary{decorations.pathmorphing}
\usepackage{breqn}
\usepackage{cancel}
\usepackage{comment}
\usepackage{amsfonts,amssymb}
\usepackage{xcolor}
\usepackage{tcolorbox}
\usepackage{url}
\usepackage{tikzsymbols}
\definecolor{mintcream}{rgb}{0.96, 1.0, 0.98}
\definecolor{champagne}{rgb}{0.97, 0.91, 0.81}
\definecolor{bubblegum}{rgb}{0.99, 0.76, 0.8}
\definecolor{airforceblue}{rgb}{0.36, 0.54, 0.66}
    	\definecolor{almond}{rgb}{0.94, 0.87, 0.8}
    	\definecolor{antiquefuchsia}{rgb}{0.57, 0.36, 0.51}
    	\definecolor{antiquewhite}{rgb}{0.98, 0.92, 0.84}
\usepackage[offset=1.2em]{simpler-wick}
\usepackage[compat=1.1.0]{tikz-feynman}
\usepackage{tikz-3dplot}
\usepackage{diagbox}

\usepackage{array}
\usepackage{arydshln}

\usepackage[pdftex,unicode,
colorlinks=true,
linkcolor = black,
citecolor = cyan,
urlcolor = airforceblue]{hyperref}
\usepackage[colorinlistoftodos]{todonotes}

\newcommand{\Tr}{\operatorname{Tr}}

\newcommand{\ev}[1]{\Big\langle #1 \Big\rangle}

\begin{document}
\hfill 

\hfill

\bigskip
\begin{center}
    \text{\Large{Superintegrability of the monomial Uglov matrix model}}
\end{center}
\bigskip

\centerline{{ V. Mishnyakov}$^{a,}$\footnote{victor.mishnyakov@su.se},
	{ I. Myakutin}$^{b,}$\footnote{miakutin.ia@phystech.edu},}

\bigskip

\begin{center}
	$^a$ {\small {\it Nordita, KTH Royal Institute of Technology and Stockholm University,}}\\{\small {\it 
Hannes Alfv\'ens v\"ag 12, SE-106 91 Stockholm, Sweden}}
\\
	$^b$ {\small {\it MIPT, Dolgoprudny, 141701, Russia}}\\
\end{center} 
       \centerline{\bf \normalsize Abstract}
        \vspace{0.3cm}
        {
        In this paper we propose a resolution to the problem of $\beta$-deforming the non-Gaussian monomial matrix models. The naive guess of substituting Schur polynomials with Jack polynomials does not work in that case, therefore we are forced to look for another basis for superintegrability. We find that the relevant symmetric functions are given by Uglov polynomials, and that the integration measure should also be deformed. The measure appears to be related to the Uglov limit as well, when the quantum parameters $(q,t)$ go to a root of unity. The degree of the root must be equal to the degree of the potential. One cannot derive these results directly, for example, by studying Virasoro constraints. Instead, we use the recently developed techniques of $W$-operators to arrive at the root of unity limit.  From the perspective of matrix models this new example demonstrates that even with a rather nontrivial integration measure one can find a superintegrability basis by studying the hidden symmetry of the moduli space of deformations.}

\newpage
\tableofcontents

\section{Introduction}

Matrix models are one of the cornerstone subjects of mathematical physics \cite{Morozov:1994hh}. They appear naturally in various contexts from statistical physics \cite{forrester2010log}, quantum gravity and string theory \cite{Kazakov:1985ea,Brezin:1990rb,Gross:1989vs} to supersymmetric localization of path integrals \cite{Pestun:2016zxk}. They are also interesting to study on their own, as they serve as a playground for methods of quantum field theory and string theory, and often also as a guiding tool for related subjects. One of the key properties of matrix models is their relation to integrability, realized in the form of the well-known slogan \emph{partition function are $\tau$-functions}. This subject is quite old and well studied \cite{Morozov:1994hh,Mironov:2002iq,Kharchev:1991gd}. On the other hand a lot of attention has been recently attracted to the phenomenon of \emph{superintegrability} or put simply \emph{exact solvability} of matrix models \cite{Mironov:2017och,Mironov:2022fsr,Mironov:2020tjf,Mironov:2022gvd,Cassia:2020uxy,Mironov:2024jdu,Wang:2022wei}. This property is manifested in the existence of explicit formulas for correlators at finite $N$ for certain types of matrix models. The statement is complementary to the ordinary integrability, i.e. it requires other hidden symmetries apart from those responsible for KP/Toda integrability of matrix models. Moreover \emph{superintegrability} extends beyond the integrable cases, therefore it can also be a key to understanding the deformation of standard integrability theory \cite{Morozov:2012dz,Gerasimov:1994ig,Mironov:2023hwf,Mironov:2023mve,Liu:2023trf,Bourgine:2023erh}.
\\

This paper is devoted to the construction of new exactly solvable matrix, or rather, eigenvalue models. They are not just interesting as a new example, but also for the kinds of structures that they are related to. For instance, to construct these new examples, we apply our intuition about hidden symmetries responsible for superintegrability. Superintegrability seems to be tied to the theory of characters and in general symmetric functions naturally labelled by partitions. In fact, it was discovered recently that the combinatorial nature of the exact solutions for the averages of these symmetric functions has an algebraic structure behind it. In particular, an important role is played by algebras which belong to the family of so-called BPS-algebras \cite{Harvey:1996gc,Li:2020rij,Galakhov:2023gjs}. The relevance of representations of the $\mathcal{W}_{1+\infty}$ algebra, the Affine Yangian $Y\left( \hat{\mathfrak{gl}}_1 \right)$ \cite{Tsymbaliuk:2014fvq,Prochazka:2015deb,Schiffmann:2012tbu,Litvinov:2020zeq} were observed extensively \cite{Mishnyakov:2022bkg,Mironov:2020pcd}. Superintegrability appears as a consequence of special representations of these algebras with states indexes by partitions. On the one hand, questions arising in matrix models can be used to establish further understanding of these algebras and motivate questions. This path was taken, for example, in \cite{Mironov:2023wga, Mironov:2023zwi} where certain commutative subalgebras of $Y\left( \hat{\mathfrak{gl}}_1 \right)$ were identified and were related to novel integrable systems. On the other hand as we demonstrate here one can borrow ideas from the relevant representation theory to extend the notion of superintegrability to new examples.

To be more specific, we remind that matrix models deal with integrals of the kind:
\begin{equation}
    \ev{f\left(\Tr X^k \right)} = \int DX \exp\Big( \Tr V(X)\Big)\, f\left(\Tr X^k \right)
\end{equation}
where the integration goes over a space of matrices of a given kind, and the $V(X)$ function is called the potential. However, in most cases, when talking about generalizations and deformations, one usually represents the model as an integral over eigenvalues. The initial integral is unitary invariant, hence one can switch to integration over eigenvalues. A quite general form of an eigenvalue models will look like:
\begin{equation}
    \ev{f(x)} = \int  \prod_{i=1}^N dx_i \prod_{i<j}\mu(x_i,x_j) \exp\left(\sum\limits_{i=1}^N V(x_i) \right) f(x) \,.
\end{equation}
Here the function $\mu(x_i,x_j)$ may, in general, depend on the eigenvalues $x_i$'s in quite a complicated manner. In matrix models and a wide class of their generalizations it is a function  of differences of eigenvalues, i.e. one has: $\prod\limits_{i<j}\mu(x_i-x_j)$. In particular for eigenvalue integrals coming from Hermitian matrix models one has $\mu(x_i,x_j) = (x_i-x_j)$. However in other cases, among which are the $(q,t)$-models, the Uglov model studied here, and, say, the Muttalib--Borodin ensembles \cite{KAMuttalib_1995,BORODIN1998704,10.1214/17-EJP62}, these functions can be more complicated.  
\\

Superintegrability is a property where, for special potentials $V(x)$ and special measures $\mu(x_i,x_j)$ one can find a basis in the space of symmetric functions such that an explicit formula for expectation values exists. Moreover, a characteristic feature of these formulas is that the answer is also represented in terms of the same functions, but evaluated at special points:
\begin{equation}
    \ev{f(x)} \sim f(*)
\end{equation}
This property is known in many examples. We will review the ones that are relevant to our result in the next section. However, for many other examples we refer to \cite{Mironov:2022fsr}. Among these examples, the simplest ones correspond to Hermitian matrix models with different potentials, where the role of special symmetric functions is played by Schur polynomials which are characters of the symmetric group. Other cases involve $\beta$ and $(q,t)$-deformations, as well as introduction of dependence on external matrices.

\paragraph{Main result.}
We would like to state the main result here and refer to the bulk of the paper for the explanation of notations. It is well known that an explicit formula exists for the monomial matrix models. Namely, in that case for Schur polynomials one has \cite{Cordova:2016jlu}:
\begin{equation}
    \begin{gathered}
                \ev{S_R(\Tr X^k)}_{s,a} := \int\limits_{C_{s,a}^{\otimes N}} D X \exp\left( -\dfrac{\Tr X^s}{s} \right) S_R(X)
        \\ 
    \ev{S_R(\Tr X^k)}_{s,a} = \left(\prod_{(i,j) \in R} [[N+j-i]]_{s,a} [[N+j-i]]_{s,0} \right)  S_R \left\{\delta_{k,s} \right\}         
    \end{gathered}
\end{equation}
The integration contours and the notation in the r.h.s. are explained in the next section.
As it well known that hermitian matrix models with various potentials survive the $\beta$-deformation \cite{Morozov:2012dz}, a natural question arises, whether it is possible to do for the monomial potential. We propose the following conjecture:
\begin{tcolorbox}[colback=mintcream!70]
\begin{equation}
\begin{gathered}
    \ev{U^{(s,1)}_R(x_i)}_{a}  =  \int\limits_{C^{\otimes N}_{s,a}} \prod_{i=1}^N dx_i \prod_{i<j}(x_i^s-x_j^s)^{\frac{2(\beta-1)}{s}}(x_i-x_j)^{2} \exp\left( - \sum\limits_{i=1}^N \dfrac{x_i^s}{s} \right) U^{(s,1)}_R(x_i) 
    \\
            \ev{U_R^{(s,1)}}_a=  \left( \prod_{(i,j) \in R}[[\beta N - \beta (i-1) +(j-1)]]_{s, 0} \cdot [[\beta N - \beta i +j]]_{s, a} \right) \cdot\frac{U_R^{(s,1)}(\delta_{k, s})}{\beta^{|R|/s}} \, .
\end{gathered}
\end{equation}
\end{tcolorbox}
We call this eigenvalue model the monomial Uglov model, following \cite{Khlaif:2022mhv} where the Uglov matrix model has been introduced. One can see that the measure $\mu(x_i,x_j)$ in this case is more complicated. The basis for superintegrability is now given by the so-called Uglov polynomials, which are a deformation of Schur polynomials, similar to Jack polynomials. They reduce, together with the measure, and, hence, the whole formula to the undeformed case at $\beta \rightarrow 1$. In the paper we also propose a second version of this formula for a slightly different measure. Let us point out a few interesting relations that this model and its components have to other subjects. 
\begin{itemize}
    \item The key component is the Uglov limit. The Uglov polynomials were originally introduced by D.Uglov in \cite{Uglov:1997ia} as a special limit of Macdonald polynomials, with $(q,t)$ around a primary root of unity point. There they were shown to be in one-to-one correspondence with the eigenfunctions of the $\mathfrak{gl}_s$ spin-Calogero model. This is interesting on its own, as the relevance of the ordinary trigonometric Calogero system for $\beta$ deformed matrix models was observed in various scenarios \cite{Mironov:2020pcd}. They also recently appeared in various other contexts related to conformal field theory and representation theory of the super-Virasoro algebra \cite{Bershtein:2022fkn,Belavin:2020gil}
    \item As it was shown recently the statement above could be further uplifted. In the case of Jack polynomials they can be shown to not only be eigenfunctions of the Calogero Hamiltonian, but actually states in the Fock representation of the full $Y\left( \mathfrak{\hat{gl}}_1 \right)$ affine Yangian.  Uglov polynomials, on the other supposedly play the same role in higher rank affine Yangians $Y\left( \mathfrak{\hat{gl}}_s \right)$ (see \cite{Galakhov:2024mbz} where rank two $s=2$ was discussed, and also  \cite{Chistyakova:2021yyd}). 
    \item The measure of the Uglov matrix model was first introduced in \cite{Kimura:2011gq}. It was motivated by considering the matrix models representation of the Nekrasov partition function on orbifolds $\mathbb{C}^2/\Gamma_{s,v}$. \footnote{We use a slightly different notation compared to \cite{Kimura:2011gq} as in our case we preferred to keep $s$ for the degree of the potential.}. As we will also stress later, our results only apply to the cases, which correspond to $\mathbb{C}^2/\mathbb{Z}_s$ and $\mathbb{C} \times \left(\mathbb{C}/\mathbb{Z}_s \right)$. The potential for the instanton matrix model is more complicated to study in contexts of superintegrability. Various discussions about the relevance of the root of unity limit were already presented \cite{Cordova:2016jlu}. We hope our works serves as a further clarification of the subject. The relevance of the root of unity limit for gauge theories on orbifolds has been demonstrated in numerous works \cite{Belavin:2011sw,Nishioka:2011jk,Itoyama:2013mca}
\end{itemize}

The way that one can come up with this formula is interesting in itself. Let us briefly summarize the logic here. First, one has to somehow come up with the correct measure. Even then, as we will explain in the paper it is not possible to extend methods like \cite{Cassia:2020uxy,Mishnyakov:2022bkg} to just derive the solution of the Uglov matrix model straightforwardly from Virasoro constraints. On the other hand, interpolating at different integer $N$ gets quite complicated  quite fast, hence without knowing the correct basis of Uglov polynomials it does not seem plausible to guess it. Hence this is a very illuminating test case for using hidden symmetries instead. The key is guessing that the root  of unity limit:
\begin{equation}
    \quad q = \omega_r u,\quad t = \omega_r u^{\beta}, \text{ and } u \to 1
\end{equation}
is relevant to the non-Gaussian model. To do this one has to turn to techniques developed in \cite{Mironov:2020pcd}. Using techniques developed there one can reverse engineer the $W$-operator, that generated the undeformed monomial model. That allows to identify it as a root of unity limit of certain element of the quantum toroidal DIM algebra \cite{Miki:2007mer,Ding:1996mq,Liu:2023trf}, with the $(q,t)$-dependent Macdonald operator being its building block. On the algebraic side, the relevance of the root of unity  limit becomes clear. Even without explicitly studying the form of the deformed $W$-operator we then conjecture the superintegrability of the Uglov model. Let us mention here, that the root of unity limit of quantum algebras is interesting on its own \cite{Itoyama:2014pca,Kimura:2019xzj} and its role in the matrix model should be studied further. The described logic can be summarized by the following picture:
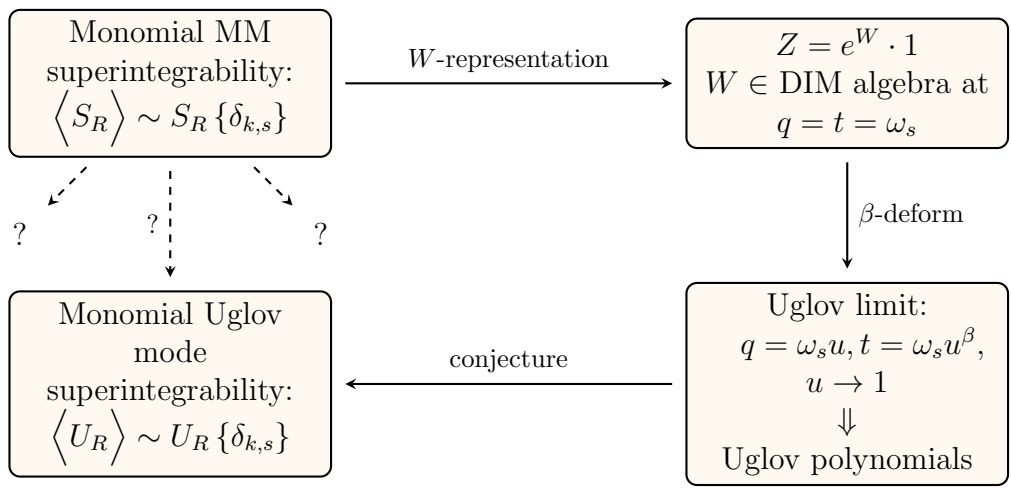
\begin{figure}[H]
    \centering
    \begin{tikzpicture}[shorten <=5pt,shorten >=5pt,>=stealth,style=thick]
         \node[fill=champagne!30,draw,rounded corners] (A) at  (-4.5,2) {\begin{minipage}{4cm}
         \centering
            Monomial MM superintegrability:\\
             $\ev{S_R} \sim S_R\left\{\delta_{k,s} \right\}$
         \end{minipage}};
         \node[fill=champagne!30,draw,rounded corners] (B) at (4.5,2) {\begin{minipage}{4cm}
         \centering
         $Z=e^{W} \cdot 1$
         \\
         $W \in $ DIM algebra at $q=t=\omega_s$
         \end{minipage}};
           \node[fill=champagne!30,draw,rounded corners] (C) at (4.5,-2) {\begin{minipage}{4cm}
         \centering
         Uglov limit:
         \\
         $\quad q = \omega_s u, t = \omega_s u^{\beta}, $
         \\
         $u \to 1$
         \\
         $\Downarrow$ 
         \\
         Uglov polynomials
         \end{minipage}};
    \node[fill=champagne!30,draw,rounded corners] (D) at (-4.5,-2) {\begin{minipage}{4cm}
         \centering
         Monomial Uglov mode superintegrability:\\
             $\ev{U_R} \sim U_R\left\{\delta_{k,s} \right\}$
         \end{minipage}};
         \draw[->] (A) -- node[above,midway] {\footnotesize{$W$-representation} } (B);
          \draw[->] (B) -- node[right,midway] {\footnotesize{$\beta$-deform} } (C);
        \draw[->] (C) -- node[above,midway] {\footnotesize{conjecture} } (D);

          \draw[->,dashed] (A) -- node[left,midway] {\footnotesize{?} } (D);
          \node (A1) at (-2.5,0) {?};
           \node (A2) at (-6.5,0) {?};
        \draw[->,dashed] (A) --  (A1);
                \draw[->,dashed] (A) --  (A2);
    \end{tikzpicture}
    \caption{Schematic of the logic that leads to superintegrability of the Uglov model.}
    \label{fig:logic}
\end{figure}
A key difficulty here is that one cannot prove this statement by most methods, that work in the undeformed case. In particular in \cite{Cordova:2016jlu} the proof utilized the technique of orthogonal polynomials, which breaks down after $\beta$-deformation, just as all the methods related to free fermions and KP/Toda integrability. Instead, below we provide  evidence for its validity that we were able to find.

\paragraph{The paper is organized as follows. }

In section \ref{sec:SMM} we review the superintegrability property of monomial matrix models \cite{Cordova:2016jlu}. We discuss contour choices and provide a few examples.

In section \ref{sec:UglovLimit} we introduce the main players of our construction: the Uglov matrix model and the Uglov polynomials. We discuss the peculiarities of the root of unity limit. In fact, the limit produces a bunch of measures, indexed by an additional parameter $v$. Whereas taking the same limit in Macdonald polynomials seems to be problematic except for two special cases $v=0$ and $v=1$.

In section \ref{sec:Result} we formulate our result in detail and also present the second formula for the $v=0$ case. We provide some examples and comment on a specific feature of the Gaussian case.

In section \ref{sec:Support} we go on to list arguments in support of our conjecture. Among those is the key motivation about $W$-operators where we explain in greater detail the logic outlined in fig. \ref{fig:logic}

\section{Superintegrability and the monomial matrix model}\label{sec:SMM}
As discussed above, a plethora of models possessing the superintegrability property are known. Among those are the series of monomial matrix model, given by the following matrix integrals:
\begin{equation}
  \ev{f\left(\Tr X^k \right)} = \int DX \, \exp\left(-\dfrac{\Tr X^s }{s}\right) f\left(\Tr X^k \right)
\end{equation}
where the integral goes over $N \times N$ matrices, however, the integration contour is still to be defined.
\\

Let us start with the simplest case, which is the Gaussian model for $s=2$. It can be treated simply by Wick's theorem and clever rewriting of contraction in terms of symmetric group characters \cite{Mironov:2017och,Corley:2001zk}. Nevertheless, superintegrability is a nontrivial statement about the special form of averages of Schur polynomials:
\begin{equation}\label{eq:GaussianSuperint}
    \Big\langle S_R(\Tr X^k) \Big\rangle  = \dfrac{S_R \left\{ N \right\} }{S_R \left\{\delta_{k,1} \right\} } S_R \left\{\delta_{k,2} \right\}= \left(\prod_{(i,j) \in R}(N+j-i)  \right) S_R \left\{\delta_{k,2} \right\}
 \end{equation}
Here we have used a notation for Schur functions evaluated at special loci of their variables, which we explain in detail in Appendix \ref{sec:AppendixNotations}. The rightmost expression represents the $N$ dependence of the expectation value as a product of simple factors over the boxes of the Young diagram $R$. The value of $(j-i)$ is called the content of a box in the Young  diagram. For that reason we will sometimes refer to such parts of formulas as the content product. Since superintegrability of matrix models is not yet a completely formalised property one can think of both the middle and the rightmost parts of the formula as its manifestation. 
\\

As announced in the introduction, it is well known that superintegrability of the Gaussian matrix model survives deformations. The $\beta$ deformation is introduced by switching to integration over eigenvalues. First, since the matrix model is unitary invariant we can integrate out the angular degrees of freedom. This will leave us with an $N$-fold integral over eigenvalues and produce a Jacobian, which in this case is given by the square of the Vandermonde determinant:
\begin{equation}
    \ev{f(x_i)} = \int \prod_{i=1}^N dx_i\,\Delta^2(x) \exp\left( - \dfrac{1}{2}\sum_{i=1}^N x_i^2\right) f(x_i)
\end{equation}
where $\Delta(x) = \prod\limits_{i<j}(x_i-x_j)$ is the Vandermonde determinant. The deformation is introduced by switching a parameter in the power of the Vandermonde determinant:
\begin{equation}
    \Delta^2(x) \rightarrow \Delta^{2\beta}(x)
\end{equation}
It is well known that under such deformation of the integration measure preservation of superintegrability requires the substitution of Schur symmetric polynomials by the $\beta$-dependent Jack symmetric polynomials. In that case one has:
\begin{equation}\label{eq:JackSuperint}
    \ev{J_R(x_i)} =\beta^{\frac{|R|}{2}} \dfrac{J_R\left\{ N \right\}}{J_R\left\{ \delta_{k,1 }\right\}} J_R\left\{ \delta_{k,2} \right\} = \beta^{- {|R| \over 2}} \left(\prod_{(i,j) \in R}\left(\beta N-\beta (i-1)+(j-1) \right)  \right) J_R\left\{ \delta_{k,2} \right\}
\end{equation}
As we can see, both manifestations of superintegrability get deformed. The content product now also depends on $\beta$. This is now a more peculiar statement, since at generic $\beta$ we cannot use Wick's theorem effectively. Moreover, lots of other techniques such as unitary Harish-Chandra–Itzykson–Zuber integration, orthogonal polynomials and others, related to free fermion representation and KP/Toda integrability of matrix models  break down. The only straightforward way to deduce this formula, would be by interpolating at multiple integer values of $\beta$. Instead, a proof of these formulas required to look at algebraic structures behind superintegrability, which in this case appears to be the affine Yangian $Y(\hat{\mathfrak{gl}}_1)$ \cite{Mironov:2020pcd,Mishnyakov:2022bkg,Wang:2022lzj} 
\\

In contrast to the Gaussian case, the monomial matrix models for $s>2$ are not uniquely defined by the potential, but also involve a contour ambiguity, i.e. they require a choice of a correct integration contour in order for the integral to be convergent. There are various peculiarities related to these contour choices, from Virasoro constraints \cite{Cassia:2021dpd} to Dijgraaf-Vafa phases, which we do not touch here \cite{Dijkgraaf:2002dh,Alexandrov:2003pj}. This question has been addressed in a specific way in \cite{Cordova:2016jlu}, where it is proved that averages of Schur function are exact for special integration contours. In this particular case the averages in the model are defined as:
\begin{equation}
    \ev{S_R(\Tr X^k)}_{s,a} := \int\limits_{C_{s,a}^{\otimes N}} D X \exp\left( -\dfrac{\Tr X^s}{s} \right) S_R(X)
\end{equation}
Here the integration contour is such that all $N$ eigenvalues are integrated over the contour $C_{s,a}$. To define it, denote by \(B_{s, j}\) the straight ray from 0 to \(\omega_s^j\cdot\infty\), where $\omega_s=e^{2\pi i \over s}$, then:
\begin{equation}
    C_{s, a} = \sum\limits_{j = 0}^{s-1} \omega_s^{-ja} B_{s,j}  , \quad a\neq0
\end{equation}
Different contours are labelled $a$, which runs from 0 to $s-1$. For example, for $s=3$ the integration contours can be drawn as:
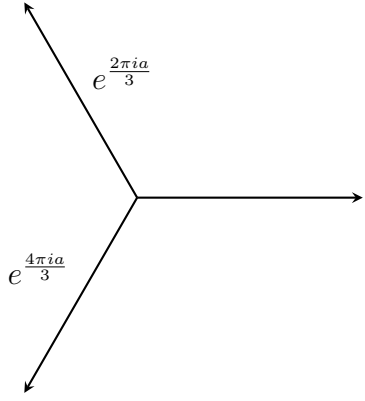
\vspace{0.2cm}
\begin{center}
    \begin{tikzpicture}[>=stealth,style=thick]
    \node (A) at (0,0) {};
    \draw [->]  (0,0) -- (0:3cm);
    \draw [->]  (0,0) --  node [midway, above right] {$ e^{\frac{2\pi i a}{3}}$}  (120:3cm);
    \draw [->]  (0,0) --  node [midway, above left] {$ e^{\frac{4\pi i a}{3}}$}  (240:3cm);
    
\end{tikzpicture}
\end{center}
\vspace{0.2cm}
Here for each $a$ the corresponding ray is multiplied by a root of unity coefficient. For $a \neq 0$ these contours are closed and therefore form a basis in the space of admissible closed contours where the integrals converge. One can perform calculations in this model by simply expanding the Vandermone and reducing the problem to single variable integrals. The contours are chosen in such a way, that single variable moments are given by:
\begin{equation}
    \int\limits_{C_{s, a}} x^ke^{-\frac{x^s}{s}}\, dx = \delta_{k+1-a}^{\left(s\right)}\int_{0}^{\infty} x^ke^{-\frac{x^s}{s}}\, dx = \delta_{k+1-a}^{\left(s\right)} \, s^{\frac{1+k-s}{s}}\Gamma \left(\frac{k+1}{s} \right)
\end{equation}
where $\delta_k^{(s)} = 1$ if $s|k$ and $\delta_k^{(s)} = 0$ else. This also means that normalized averages are defined only for $N=0 \mod s$ and $N=a \mod s$, when $\ev{1} \neq 0$. The case of non-normalized averages was studied in \cite{Barseghyan:2022txq} where rectangular Schur polynomials play a distinguished role.
\\

Finally, for each choice of contour and proper choice of $N$, the exact expression for averages of Schur functions is given by \cite{Cordova:2016jlu}:
\begin{equation}\label{eq:MonomialSuperint}
    \ev{S_R(\Tr X^k)}_{s,a} = \left(\prod_{(i,j) \in R} [[N+j-i]]_{s,a} [[N+j-i]]_{s,0} \right)  S_R \left\{\delta_{k,s} \right\} 
\end{equation}
where
\begin{equation}\label{eq:doublebracket}
    [[n]]_{s,a} = \left\{ \begin{split}
        &n \, , \, n=a  \mod s
        \\
        &1 \, , \, n \neq a  \mod s
    \end{split} \right.
\end{equation}
This expression essentially can be understood as a generalization of \eqref{eq:GaussianSuperint}, where now the product is taken not over all boxes of the partition but only over certain diagonals. Note that each of the factors in \eqref{eq:MonomialSuperint} also depends on $N \mod s$. Take $s=3$ as an example again and suppose $N = 0 \mod 3$. The for the partition $[5,4,3,2,2]$ one has:
\begin{center}
		\begin{equation*}
  \begin{split}
      		[[N+j-i]]_{3,0} \, &: \,		\ytableausetup{boxsize=1em,aligntableaux = center}  \quad \ydiagram{5,4,2,2,2}*[*(almond) ]{1,1+1,2+1,1,1+1}*[*(almond) ]{3+1} \quad  \rightarrow \quad N^3(N+3)(N-3)^2
	\\
 \\
		[[N+j-i]]_{3,1} \, &: \,		\ytableausetup{boxsize=1em,aligntableaux = center}  \quad \ydiagram{5,4,2,2,2}*[*(antiquefuchsia) ]{1+1,2+1,1,1+1}*[*(antiquefuchsia) ]{4+1} \quad \rightarrow \quad (N+1)^2(N+4)(N-2)^2
  \end{split}
	\end{equation*}
	\end{center}
 Where the colored boxes are the ones that enter the content product. As visible in the example, these are located along the diagonals that are shifted three boxes apart from each other. At the same time for $N=1 \mod 3$ one has:
\begin{center}
		\begin{equation*}
  \begin{split}
		\hspace{-0.96cm} [[N+j-i]]_{3,1} \, &: \,		\ytableausetup{boxsize=1em,aligntableaux = center}  \quad \ydiagram{5,4,2,2,2}*[*(antiquefuchsia) ]{1,1+1,2+1,1,1+1}*[*(antiquefuchsia) ]{3+1} \quad  \rightarrow \quad N^3(N+3)(N-3)^2
  \end{split}
	\end{equation*}
	\end{center}
Another feature of the result \eqref{eq:MonomialSuperint} is a lack of representation of the content product in terms of special values of Schur functions. As announced in the introduction, we will propose such formula below. Finally, to see how it reduces to the Gaussian case, one has to simply put $s=2$ and $a=1$. In this case, for each box the entry is either even or odd, hence  either only the first or only the second factor is non-trivial. In that manner, we obtain the product \eqref{eq:GaussianSuperint}.
\section{The Uglov limit and the Uglov matrix model}\label{sec:UglovLimit}
As we will explicitly demonstrate below, the naive $\beta$ deformation with substitution of the Vandermonde and Schur function for Jack polynomials does not work. Instead, we should introduce a new measure and a new set of symmetric functions. In this section we describe both in detail and explain the motivation later. In both cases they can be identified as a certain limit of the respective objects depending on two parameters $(q,t)$ related to the quantum toroidal algebra. In this limit both parameters go to a root of unity with a certain scaling:
\begin{equation}\label{eq:Limit}
    q = \omega_r u,\quad t = (\omega_r u)^{\beta}, \text{ and } u \to 1
\end{equation}
As we will see the limiting procedure strongly depends on the $v = \beta \mod r$ and in fact we will mostly focus in two cases:
\begin{equation}
     \beta \!\! \mod r = 1  \quad  \Rightarrow \quad q = \omega_r u,\quad t = \omega_r u^{\beta}, \text{ and } u \to 1
\end{equation}
which is the limit that was  considered by Uglov \cite{Uglov:1997ia}. And 
\begin{equation}
     \beta \!\! \mod r = 0  \quad  \Rightarrow \quad q = \omega_r u,\quad t =  u^{\beta}, \text{ and } u \to 1
\end{equation}

\subsection{The Uglov Matrix model and root of unity limit of the $(q,t)$-matrix model}

The corresponding matrix model is the so called Uglov matrix model. Just as the $\beta$-deformation it is defined in terms of the eigenvalue representation:
\begin{equation}
    \ev{f(x_i)}^{(r,v)} = \int \prod_{i=1}^N dx_i \, \Delta_{\text{Uglov}}^{(r,v)}(x) \exp\left( - \sum_{i=1}^n  V(x_i) \right) f(x_i)
\end{equation}
And hence as a deformation of the ordinary Hermitian matrix model it corresponds to the substitution:
\begin{equation}
    \Delta^2(x) \, \rightarrow \,   \Delta_{\text{Uglov}}^{(r,v)}(x) := \prod_{i<j}^N (x_i^r-x_j^r)^{\frac{2(\beta-v)}{r}} \prod_{k=0}^{v-1} (x_i-\omega_r^k x_j)^2
\end{equation}
For brevity we will call this measure an Uglov determinant by analogy with the Vandermonde determinant. As announced above, we will deal with two cases $v=0, v=1$ when:
\begin{equation}
\begin{split}
        \Delta_{\text{Uglov}}^{(r,1)}(x)& = \prod_{i<j}^N (x_i^r-x_j^r)^{\frac{2(\beta-1)}{r}}  (x_i- x_j)^2
        \\
        \Delta_{\text{Uglov}}^{(r,0)}(x)& = \prod_{i<j}^N (x_i^r-x_j^r)^{\frac{2\beta}{r}} 
\end{split}
\end{equation}
There are two special limiting cases $\beta=1$ and $r=1$. First, at $\beta=1$ the $\Delta_{\text{Uglov}}^{(r,1)}(x)$ measure goes to the square of the Vandermonde determinant, which makes this deformation consistent with the undeformed case:
\begin{equation}
    \Delta^{(r,1)}_{\text{Uglov}}(x)  \, \xrightarrow{\quad \beta \, \rightarrow  \, 1 \quad  } \, \Delta^2(x)
\end{equation}
On the other hand at $r=1$ we obtain the standard $\beta$ deformed Vandermonde:
\begin{equation}
    \Delta_{\text{Uglov}}^{(1,v)} = \Delta^{2\beta}(x)
\end{equation}
The origin of this measure as suggested by the limit \eqref{eq:Limit} should be related to the  $\left(q, t\right)$-deformed matrix model. We will not deal with the whole $(q,t)$-models right now, but only with the measure part. Here we will briefly recall the derivation, as it was already done in \cite{Khlaif:2022mhv}. The $(q,t)$-models are defined by substituting the  Vandermonde determinant by:
\begin{equation}
    \Delta^2\left(x\right) \to \prod_{i \neq j}\frac{\left(\frac{x_i}{x_j}, q\right)_{\infty}}{\left(t\frac{x_i}{x_j}, q\right)_{\infty}}\cdot \prod_i x_i^{\beta(N-1)}
\end{equation}
Where $\left(z, q\right)_{\infty} = \prod_{i=0}^{\infty}\left(1 - zq^i\right)$ - is the Pochhammer symbol. The integral is then given either by the Jackson integral or a special contour, reproducing the Jackson sum.  Consider the limit \eqref{eq:Limit}, with $\beta =k \cdot r +v$:
\begin{equation}
    q = \omega_r u,\quad t = \omega_r^v u^{\beta}, \text{ and } u \to 1
\end{equation}
Under this limit the $(q,t)$ measure behaves as follows (for generic $v$):
\begin{equation}
    \frac{\left(\frac{x_i}{x_j}, q\right)_{\infty}}{\left(t\frac{x_i}{x_j}, q\right)_{\infty}} = \frac{\left(\frac{x_i}{x_j},\omega_ru\right)_{\infty}}{\left(\omega_r^vu^{\beta}\frac{x_i}{x_j}, \omega_ru\right)_{\infty}}=\frac{\left(\frac{x_i}{x_j},\omega_ru\right)_{\infty}}{\left(\omega_r^\beta u^{\beta}\frac{x_i}{x_j}, \omega_ru\right)_{\infty}}=\prod_{l=0}^{\beta -1} \left(1 - \left(\omega_ru\right)^l \frac{x_i}{x_j}\right)
\end{equation}
The crucial part here is that we used the identity: $\omega_r^v = \omega_r^\beta$, which is only true if $\beta = v \mod r$. Because of that we obtain different expressions for the measure for different remainders of division $\beta$ by $r$.
After taking the limit  \eqref{eq:Limit} we obtain the following form of measure:
\begin{equation}
    \prod_{l=0}^{\beta -1} \left(1 - \left(\omega_ru\right)^l \frac{x_i}{x_j}\right) \to \left(1-\frac{x_i^r}{x_j^r}\right)^k\prod_{l= k\cdot r}^{\beta -1} \left(1 - \omega_r^l \frac{x_i}{x_j}\right) =  \left(1-\frac{x_i^r}{x_j^r}\right)^k\prod_{l= 0}^{v-1} \left(1 - \omega_r^l \frac{x_i}{x_j}\right)
\end{equation}
Here we used that $\omega_r^{l_1} = \omega_r^{l_2}$ if $l_1 = l_2 \mod r$, and that $1-a^r = \prod_{i=1}^r (1-\omega_r^ia)$.
For generic $v$ the measure has complex coefficients. For $v=0,1$, however, is does not. For $v=1$ we obtain:
\begin{equation}
    \left(1-\frac{x_i^r}{x_j^r}\right)^k\prod_{l= 0}^{0} \left(1 - \omega_r^l \frac{x_i}{x_j}\right) = \left(1-\frac{x_i^r}{x_j^r}\right)^k \left(1 - \frac{x_i}{x_j}\right)
\end{equation}
For $v = 0$:
\begin{equation}
    \left(1-\frac{x_i^r}{x_j^r}\right)^k\prod_{l= 0}^{-1} \left(1 - \omega_r^l \frac{x_i}{x_j}\right) = \left(1-\frac{x_i^r}{x_j^r}\right)^k 
\end{equation}
\subsection{Uglov polynomials}
The key feature of Uglov limit, which was crucial for \cite{Uglov:1997ia} is that it is consistent and well-defined for Macdonald polynomials. 
\\\\
Denote the Macdonald polynomials as $\text{Mac}_R(p_k|q,t)$, depending on the integer partition $R$. Now we want to define the so-called Uglov polynomials as the limit  \eqref{eq:Limit} of Macdonald polynomials. Here we are forced to consider $v=0$ or $v=1$ as only in these cases the limit seems to be well-defined. It is interesting to study whether extensions to other values of $v$ are possible. Hence, we define:
\begin{equation}
    U^{(r,v)}_R(p_k \, | \,\beta) :=  \lim_{u \rightarrow 1} \text{Mac}_R(p_k \, | \, \omega_r u ,\omega_r^{v} u^\beta) \, ,\quad v=0,1 
\end{equation}
In the paper we drop the notation, specifying the dependence  on $\beta$ and just denote:
\begin{equation}
    U_R^{(r,v)}(p_k) := U^{(r,v)}_R(p_k \,|\, \beta) 
\end{equation}
wherever the value of $\beta$ is not important. There are several comments to make here. First, at $v=1$ we could have set $q=\omega_r^l u, q=\omega_r^l s^{\beta}$. In this case, the resulting polynomial does not depend on $l$. For $r=1$ these polynomials are nothing but Jack polynomials:
\begin{equation}
 U_R^{(1,0)}(p_k)=J_R(p_k)  
\end{equation}
For $v=1$ it makes sense to take the $\beta \rightarrow 1$ limit. In this case $q=t$ and it is well known that Macdonald polynomials reduce to Schur polynomials at that locus for any values of $q=t$, hence this is also true for Uglov polynomials:
\begin{equation}
    U_R^{(r,1)}(p_k \, |\, \beta= 1)   = S_R(p_k) 
\end{equation}
Table \ref{tab:Uglov1} presents some examples up to level 3. 
\begin{table}[htb]
\begin{center}
\ytableausetup{boxsize=0.5em,aligntableaux = center} 
	\begingroup
	\centering
	\renewcommand{\arraystretch}{2.5}
	\newcolumntype{M}[1]{>{\centering\arraybackslash}m{#1}}
\begin{tikzpicture}
	\node (table) [inner sep=0pt] {
		\begin{tabular}{M{2.185cm}|M{3cm}|M{3cm}|M{3.9cm}|M{3.4cm}}
			   \backslashbox{\hspace{0.4cm}$R$}{ \hspace{0.55cm}\vspace{-0.5cm}$r$ } & $r=1$ & $r=2$ &$r=3$&$r = 4$\\ 
                \hline
                \ydiagram{1} & $p_1$ & $p_1$ & $p_1$&$p_1$
                \\
                \hline
                 \ydiagram{2} & $\frac{p_1^2}{2}-\frac{p_2}{2}$ & $\frac{p_1^2}{2}-\frac{p_2}{2}$&$\frac{p_1^2}{2}-\frac{p_2}{2}$ & $\frac{p_1^2}{2}-\frac{p_2}{2}$ \\

                 \hline
                 \ydiagram{1, 1} &$\frac{\beta p_1^2+p_2}{\beta+1}$ &$\frac{\beta p_2+p_1^2}{\beta+1}$ &$\frac{1}{2} \left(p_1^2+p_2\right)$&$\frac{1}{2} \left(p_1^2+p_2\right)$ \\
                 \hline
                 \ydiagram{3} & $\frac{p_1^3}{6}-\frac{p_2 p_1}{2}+\frac{p_3}{3}$ & $\frac{p_1^3}{6}-\frac{p_2 p_1}{2}+\frac{p_3}{3}$&$\frac{p_1^3}{6}-\frac{p_2 p_1}{2}+\frac{p_3}{3}$ & $\frac{p_1^3}{6}-\frac{p_2 p_1}{2}+\frac{p_3}{3}$\\
                 \hline
                \ydiagram{2,1} & $\frac{\beta p_1^3-(\beta-1) p_2 p_1-p_3}{2 \beta+1}$ & $\frac{1}{3} \left(p_1^3-p_3\right)$& $\frac{((\beta+1) p_1^3+(\beta-1) p_2 p_1-2 \beta p_3)}{2(1+2 \beta)}$ &$\frac{1}{3} \left(p_1^3-p_3\right)$\\
                \hline
                \ydiagram{1,1,1} & $\frac{\beta^2 p_1^3+3 \beta p_2 p_1+2 p_3}{\beta^2+3 \beta+2}$ & $\frac{3 \beta p_2 p_1+p_1^3+2 p_3}{3 \beta+3}$&$\frac{2 \beta p_3+p_1^3+3 p_2 p_1}{2(2+\beta)}$& $\frac{\left(p_1^3+3 p_2 p_1+2 p_3\right)}{6} $
		\end{tabular}
	};
	\draw [rounded corners=.2em,line width=0.45mm] (table.north west) rectangle (table.south east);

\end{tikzpicture}
\endgroup
\end{center}
\caption{Examples of Uglov polynomials} 
\label{tab:Uglov1} 
\end{table}

The listed Uglov polynomials for $v=1$ don't look that different from each other and Jack polynomials. It becomes more visible for bigger diagrams, and, what is more important, is that properties of these functions are very different. We list some more examples in Appendix \ref{sec:AppendixUglov}. Some properties of Uglov polynomials as symmetric functions are collected in the Appendix in \cite{Galakhov:2024mbz} and some will be revealed below.
\\

\paragraph{Comment on the $v=0$ polynomials.}
When $v=0$ we get different polynomials, with a couple examples provided in table \ref{tab:Uglov0}.
\begin{table}[htb]
  \begin{center}
\ytableausetup{boxsize=0.5em,aligntableaux = center} 
	\begingroup
	\centering
	\renewcommand{\arraystretch}{2} 
	\newcolumntype{M}[1]{>{\centering\arraybackslash}m{#1}}
\begin{tikzpicture}
	\node (table) [inner sep=0pt] {
		\begin{tabular}{M{2cm}|M{3.5cm}|M{3.5cm}|M{3.5cm}}
			   \backslashbox{\hspace{0.4cm}$R$}{ \hspace{0.45cm}\vspace{-0.5cm}$r$ } & $r=2$ & $r=3$ &$r=4$\\ 
                \hline
                \ydiagram{1} & $p_1$ & $p_1$ & $p_1$
                \\
                \hline
                 \ydiagram{2} & $\frac{p_1^2}{2}-\frac{p_2}{2}$ & $\frac{p_1^2}{2}-\frac{p_2}{2}$&$\frac{p_1^2}{2}-\frac{p_2}{2}$ \\

                 \hline
                 \ydiagram{1, 1} &$p_2$ &$p_2$ &$p_2$ \\
                 \hline
                 \ydiagram{3} & $\frac{p_1^3}{6}-\frac{p_2 p_1}{2}+\frac{p_3}{3}$ & $\frac{p_1^3}{6}-\frac{p_2 p_1}{2}+\frac{p_3}{3}$&$\frac{p_1^3}{6}-\frac{p_2 p_1}{2}+\frac{p_3}{3}$\\
                 \hline
                \ydiagram{2,1} & $p_1 p_2-p_3$ & $p_1 p_2-p_3$ &$p_1 p_2-p_3$\\
                \hline
                \ydiagram{1,1,1} & $\frac{\beta p_1 p_2+2 p_3}{b+2}$ & $p_3$&$p_3$
		\end{tabular}
	};
	\draw [rounded corners=.2em,line width=0.45mm] (table.north west) rectangle (table.south east);

\end{tikzpicture}
\endgroup
\end{center}
\caption{The $v=0$ analogs of Uglov polynomials} 
\label{tab:Uglov0} 
\end{table}
After looking at this table and a number of other examples one can notice that for diagrams which are divisible by $r$, i.e. such that all their row lengths are divisible by $r$, the respective polynomials only depend on power sums $p_{kr}$. Moreover one can go even further and conjecture the following formula:
\begin{equation}\label{eq:v0UglovtoJack}
    U^{(r,0)}_{r\cdot R}\left( p_{k}\, | \, r\cdot \beta \right) =  J_R\left( p_{rk}  \right)
\end{equation}
where $r\cdot R$ is the partition with all row lengths multiplied by $r$, i.e. $[r R_1, r R_2 ,\ldots ]$. We will see below how this statement interacts nicely with the Cauchy formula and later superintegrability. For example, one can take $R=[2,1]$ and $r=2$:
\begin{equation}
\begin{split}
        U_{[4,2]}^{(2,0)}(p_k|\beta)&=\frac{\beta  p_2^3-(\beta -2) p_4 p_2-2 p_6}{2 (\beta +1)}
        \\
        J_{[2,1]}(p_k|\beta)&=\frac{\beta  p_1^3-(\beta -1) p_2 p_1-p_3}{2 \beta +1}
\end{split}
\end{equation}
\paragraph{Cauchy formula}
An important formula in the theory of symmetric functions that we will use below is the so-called Cauchy formula. For Schur polynomials, it has the standard form \cite{macdonald1998symmetric}:
\begin{equation}\label{eq:Cauchy}
    \exp\left({\sum\limits_{k=1}^{\infty }\dfrac{p_k \bar{p}_k}{k} }\right) = \sum_R S_R(p_k) S_R(\bar{p_k})
\end{equation}
where $p_k$ and $\bar{p}_k$ are two independent sets of variables. Its $\beta$-deformation for Jack polynomials and Macdonald is well-known \cite{macdonald1998symmetric}. 
For the Macdonald polynomials one has:
\begin{equation}
    \exp\left({\sum\limits_{k=1}^{\infty }\frac{t^{-k/2}-t^{k/2}}{q^{-k/2}-q^{k/2}}\dfrac{p_k \bar{p}_k}{k} }\right) = \sum_R \frac{\text{Mac}_R(p_k|q,t) \text{Mac}_R(\bar{p}_k|q,t)}{\|\text{Mac}_R(p_k|q,t)\|^2}
\end{equation}
Where the normalization is given by:
\begin{equation}
    \|\text{Mac}_R(p_k|q,t)\|^2 = \prod_{(i, j) \in R} \frac{t^{(-R_j^T+i)/2}q^{(R_i -j+1)/2}+t^{(R_j^T-i)/2}q^{(-R_i +j-1)/2}}{t^{(-R_j^T+i-1)/2}q^{(R_i -j)/2}+t^{(R_j^T-i+1)/2}q^{(-R_i +j)/2}}
\end{equation}
In the Uglov limit \eqref{eq:Limit} for $v=1$ it becomes:
\begin{equation}\label{eq:CauchyUglov}
    \exp\left({\sum\limits_{k=1}^{\infty }\beta^{\delta_{k| r}}\dfrac{p_k \bar{p}_k}{k} }\right) = \sum_R \frac{U^{(r, 1)}_R(p_k) U^{(r, 1)}_R(\bar{p}_k)}{\|U_R^{r, 1}\|^2}
\end{equation}
Here the coefficient in front of power sums in the l.h.s. is $\beta$ for $k$ divisible by $r$ and one otherwise. The norm of Uglov polynomials can be computed directly from the limit.
\\

Let us go into detail here to demonstrate once again how the dependence on $\beta \mod r$ arises when taking the limit.
In particular, if we denote by $\eta_k$ - the coefficient in front of $\frac{p_kq_k}{k}$ in the l.h.s. of the Cauchy formula we have at $v=1$:
\begin{equation}
\eta_k = \frac{t^{-k/2}-t^{k/2}}{q^{-k/2}-q^{k/2}} = \frac{(\omega_r (1+h)^\beta)^{-k/2}-(\omega_r(1+h)^\beta)^{k/2}}{(\omega_r(1+h))^{-k/2}-(\omega_r(1+h)^{k/2}}    
\end{equation}
where we parameterized $u$ from  \eqref{eq:Limit} as $u=e^{h}$, where $h \to 0$. Expanding explicitly, one has:
\begin{equation}
\left(\omega_r \left(1+\epsilon\right)^\beta\right)^{-k/2}-\left(\omega_r\left(1+\epsilon\right)^\beta\right)^{k/2} = \omega_r^{-\frac{k}{2}}\left(1-\frac{\beta k}{2}\epsilon\right) -\omega_r^{\frac{k}{2}}\left(1+\frac{\beta k}{2}\epsilon\right)    
\end{equation}
thus if $ \omega_r^{k} \neq 1 $
 \begin{equation}
    \lim_{h \rightarrow 0 } \eta_{k}  = 1 \, ,
 \end{equation}
otherwise
 \begin{equation}
    \lim_{h \rightarrow 0 } \eta_{k}  = \beta \ .
 \end{equation}
This is the result above. It's also possible to take the limit in Cauchy formula for $v\neq 1$, however in these cases interesting phenomena can occur. The relevant case for us is $v=0$. After similar manipulations we arrive at:
\begin{equation}
\lim_{h\rightarrow 0 } \eta_k =  \left\{ \begin{split}
    & \frac{\beta}{\omega_r^{k/2}} = \beta \cdot (-1)^{1+\delta_{k|2r}}  \,, k = 0 \mod r
    \\
    & 0 \,, k \neq 0 \mod r
\end{split}
\right.
\end{equation}
Hence the generating function depends only on powers sums with indices that are divisible by $r$. As we see, this agrees nicely with the fact that in this case the symmetric polynomials $U_R^{(r,0)}$ with diagrams that are divisible by $r$ only depend on $p_{kr}$.
\section{Superintegrability of the non-Gaussian monomial Uglov matrix model(s)}\label{sec:Result}
\subsection{The main statement}
Finally, we have all the ingredients to explain the main statement of this paper. The Uglov matrix model can be defined for any potential. However, it appears that in this model, for superintegrability to hold, the potential, the measure and the symmetric function should all be in harmony. Namely, we put:
\begin{equation}
    r=s
\end{equation}
and consider the following averages
\begin{equation}
    \ev{U^{(s,v)}_R(x_i)}^{(s,v)}_{(s,a)}  =  \int\limits_{C^{\otimes N}_{s,a}} \prod_{i=1}^N dx_i \prod_{i<j}(x_i^s-x_j^s)^{\frac{2(\beta-v)}{s}}(x_i-x_j)^{2v} \exp\left( - \sum\limits_{i=1}^N \dfrac{x_i^s}{s} \right) U^{(s,v)}_R(x_i) 
\end{equation}
Which we shortly denote by:
\begin{equation}
    \ev{U_R^{(s,v)}}_a := \ev{U^{(s,v)}_R(x_i)}^{(s,v)}_{(s,a)}
\end{equation}
We conjecture the following superintegrability properties (separately for $v=0$ and $v=1$):
\begin{equation}\label{eq:result2}
    \begin{split}
        &v=1:
        \\
        &
        \ev{U_R^{(s,1)}}_a=  \left( \prod_{(i,j) \in R}[[\beta N - \beta (i-1) +(j-1)]]_{s, 0} \cdot [[\beta N - \beta i +j]]_{s, a} \right) \cdot\frac{U_R^{(s,1)}(\delta_{k, s})}{\beta^{|R|/s}} \, ,
        \\
        &v=0:\text{if }  R_i \,| \,s \,  ,  \,  \forall \, i \, ,
        \\
        &
        \ev{U_R^{(s,0)}}_a=   
        \begin{split}    
        \left( \prod_{\substack{(i,j) \in R \\ j = 1\mod s}}(\beta N - \beta (i-1) +(j-1)) \cdot (\beta N - \beta i +j) \right) \cdot\frac{U_R^{(s,0)}(\delta_{k, s})}{\beta^{|R|/s}}  \, , \quad  
        \end{split}
    \end{split} 
\end{equation}
Here, for $v=0$, if there is at least one row in the diagram such that $R_i$ is not divisible by $s$, the expectation value vanishes. The content product for $v=0$ may be understood in the following way: let the diagram corresponding to the polynomial be divided into rectangles of $1\cdot r$, then for every rectangle we have a single content factor in the r.h.s. 
The $\beta$ dependent content product functions $[[x]]_{s,a}$ are now understood as follows:
\begin{equation}
    [[\beta N - \beta x + y]]_{s,a} = \left\{ \begin{split}
        &\beta N - \beta x + y \, ,\quad \, N-x+y=a  \mod s
        \\
        &1 \, , \, \quad N-x+y \neq a  \mod s
    \end{split} \right.
\end{equation}
The prescription means that the product is taken over the same boxes as in the undeformed case, while the content function is deformed. Notice that the deformation is slightly different in the two brackets. It's evident that for $v=1$ and $\beta=1$ we get the original undeformed formula \eqref{eq:MonomialSuperint}.
\\\\
For example, let us take the cubic model with $v=1$ and the contour with $a=1$ and $N = 0 \mod 3$. Then for the partition $[2,2,2]$ we have:
\begin{equation}
\begin{split}
    [[\beta N - \beta (i-1) +(j-1)]]_{3, 0}  \quad &\longrightarrow  \quad N\beta (1-\beta+N\beta)
    \\
    [[\beta N - \beta i +j]]_{3,1}  \quad &\longrightarrow  \quad (\beta  N+1-3 \beta) (\beta  N+2-\beta)
    \\
    U_{[2,2,2]}^{(3,1)}\left\{ \delta_{k,3} \right\} \quad \  &= \ \quad  \frac{\beta }{6 \beta +3}
\end{split}
\end{equation}
and
\begin{equation}
    \ev{U_{[2,2,2]}^{(3,1)}}_1 = N\beta (1-\beta+N\beta)(\beta  N+1-3 \beta) (\beta  N+2-\beta)\cdot\frac{\beta }{6 \beta +3}\cdot \frac{1}{\beta^2}
\end{equation}
In the $v = 0$ case an example computation is:
\begin{equation}
\begin{gathered}
        U_{[3, 3]}^{(3,0)}\left\{ \delta_{k,3} \right\} \quad \  = \ \quad  \frac{1}{2}
    \\
    \ev{U_{[3,3]}^{(3,0)}}_1 = \beta N(\beta N - \beta +1)(\beta N - \beta )(\beta N - 2\beta +1)\frac{1}{2\beta^2}
\end{gathered}   
\end{equation}

\paragraph{Two superintegrabilities in the Gaussian model.} We would like to briefly discuss the Gaussian case separately, as it is the simplest example, and also involves an interesting feature. Notice that at $s=r=2$ the result for $v=1$ \eqref{eq:result2} expresses superintegrability of Uglov polynomials $U^{(2,1)}_R$ in the model given by:
\begin{equation}
    \prod_{i<j} (x_i^2-x_j^2)^{(\beta-1)\over 2} (x_i-x_j)^2 \exp\left(-\dfrac{1}{2} \sum\limits_{i=1}^N x_i^2 \right)
\end{equation}
However, we know that at the same time the Gaussian weight can be integrated with the standard $\beta$-deformed Vandermonde and with a Jack polynomial insertion. In terms of the Uglov model, this corresponds to taking $r=1$ and $s=2$. This means that there are two exactly solvable eigenvalue models available with the Gaussian potential. Consider for example the two expression for $R=[2]$, where formulas are very similar, and take $N=1 \mod 2$:
\begin{equation}
\begin{split}
       \int \prod_{i=1}^N dx_i \   \Delta^{2\beta}(x) \exp{\left(-\dfrac{1}{2}\sum\limits_{i=1}^N x_i^2 \right)}\dfrac{\left(\sum\limits_{i=1}^N x_i^2+ \beta \left( \sum\limits_{i=1}^N x_i \right)^{2} \right)}{{\beta+1}} & =   \dfrac{\beta N (\beta  N+1)}{\beta  (\beta +1)}
       \\
          \int\limits_{C_{2,1}^{\otimes N}} \prod_{i=1}^N dx_i \ \Delta_{\text{Uglov}}^{(2,1)}(x) \exp{\left(-\dfrac{1}{2}\sum\limits_{i=1}^N x_i^2 \right)} \dfrac{\left( \beta \sum\limits_{i=1}^N x_i^2+ \left( \sum\limits_{i=1}^N x_i \right)^{2} \right)}{{\beta+1}} & =  \frac{(\beta  N+1) (\beta (N-1)+1)}{\beta +1}
\end{split}
\end{equation}

\paragraph{Absence of superintegrability at  $r \neq s$.}
Apart from the Gaussian case,  we could not find any superintegrability property when $r\neq s$, at least with Uglov polynomials as symmetric functions. We assume the following simple check. If the model is superintegrable we expect the averages to look like  $\ev{U_R} = {P(N, \beta) \over Q(\beta)}$, where $P$ - is  a polynomial of $N$ and $\beta$, $Q$ - is a polynomial only of $\beta$ and moreover we expect both to be factorized in $N$ and $\beta$. Moreover, suppose the Uglov polynomials were not the right basis in this case. Then if there exists some other basis with superintegrable averages of the described form, then one could generically expect any average, being expended into a superintegrable basis to be of similar form. With that in mind, we evaluate expectation values of various types, say, of $\ev{U_R^{(s,1)} }\, ,\, \ev{U^{(r,1)}_R}$ of just powers sums $\ev{p_\Delta}$. In all cases we find that, for example, the denominator increases much faster than expected and  contains large prime numbers as factors, which clearly cannot appear from simple factorized polynomials in $\beta$.

Still, these are indirect arguments, that illustrate that naive attempts to extend superintegrability beyond $r=s$ were unsuccessful.

\subsection{Representation of the content product in terms of Uglov polynomials}\label{sec:ContentProduct}
    As announced in the introduction, another result of this paper is the representation of \eqref{eq:result2} in terms of Uglov polynomials evaluated at a special locus. This is an important part of the superintegrability relation. In other known cases of superintegrability in matrix models, the whole answer is expressed through special values of symmetric functions. A similar effect was not known, even for the undeformed non-gaussian monomial model. Here we conjecture an answer  for $\beta$-deformed contents in terms of Uglov polynomials. Its analog for Schur polynomials immediately follows by setting $\beta=1$ in the formulas below and is a particular case of known specializations formulas. We believe that the respective formulas for Uglov polynomials can be proven starting directly from the Macdonald polynomials, but we don't do it in this paper.
    \\
    
    The product depends not only on $a$ but also on $N \mod s$. In fact, it depends only the difference $N-a \mod s$. Therefore, it makes sense to choose a representative with $a=0$. Formulas with other $a$ can be obtained from it. Here we propose the following formula:
\begin{equation}\label{eq:BoxAsUglov}
        \left( \prod_{(i,j) \in R}[[\beta N - \beta (i-1)+(j-1)]]_{s,0} \right) U^{(s,1)}_R\left\{ \delta_{k,s} \right\} = \beta^{|R| \over s}  U^{(s,1)}_R \left( p_k = \sum_{i=1}^N \omega_s^{(i-1)k}\right)
    \end{equation}
Both parts of this formula depend on $N \mod s$. 
\\\\
Let us demonstrate how this works in a few examples.
\begin{enumerate}
    \item  First take $N = 0 \mod s $. Then:
    \begin{equation}
         p_k = \sum_{i=1}^N \omega_s^{(i-1)k}  = \delta_{k|s}N = \left\{ 
         \begin{split}
              &N \, , \, k = 0 \mod s
         \\
         & 0 \,,\, k \neq 0 \mod s
         \end{split}
         \right.
    \end{equation}
\begin{equation}
    \left( \prod_{(i,j) \in R}[[\beta N - \beta (i-1)+(j-1)]]_{s,0} \right) U^{(s,1)}_R\left\{ \delta_{k,s} \right\}  = \beta^{|R|\over s} U^{(s,1)}_R \left( p_k = \delta_{k|s} N \right) 
\end{equation}
This can be used to obtain expressions for cases, when $N =a \mod s$, i.e.:
\begin{equation}
 \left( \prod_{(i,j) \in R}[[\beta N - \beta (i-1)+(j-1)]]_{s,a} \right) U^{(s,1)}_R\left\{ \delta_{k,s} \right\}= \beta^{|R|\over s} U^{(s,1)}_R \left( p_k = \delta_{k|s} N \right)  
\end{equation}
\item When $N \neq a \mod s$ formula \eqref{eq:BoxAsUglov} still holds. To demonstrate how it works consider $s=3$. Once again set $a= 0 \mod 3$, then if $N=1 \mod 3$ :
    \begin{equation}
    p_k = \sum_{i=1}^{N} \omega_e^{(i-1)k} = \left\{ \begin{split}
         &N \, , \, k=0 \mod 3
         \\
         & 1 \, , \, k=1,2 \mod 3
    \end{split}
    \right. =    N \delta_{k|3}  + \delta_{k-1|3} + \delta_{k-2|3} 
\end{equation}
where we used the notation:
\begin{equation}
    \delta_{k|s} = \left\{ \begin{split}
         &1 , \, k=0 \mod s
         \\
         &0 \, , \, k \neq 0 \mod s
    \end{split}
    \right. 
\end{equation}
Hence, we have for $N=1 \mod s$:
\begin{equation}
\begin{split}
        \left( \prod_{(i,j) \in R}[[\beta N - \beta (i-1)+(j-1)]]_{s,0} \right) &U^{(s,1)}_R\left\{ \delta_{k,s} \right\}  = 
        \\
        &=\beta^{|R|\over s} U^{(s,1)}_R \left( p_k =  N \delta_{k|3}  + \delta_{k-1|3} + \delta_{k-2|3}  \right) 
\end{split}
\end{equation}
and, again, we have the same expression, when $N=a+1 \mod s$:
\begin{equation}
\begin{split}
       \left( \prod_{(i,j) \in R}[[\beta N - \beta (i-1)+(j-1)]]_{s,a} \right) &U^{(s,1)}_R\left\{ \delta_{k,s} \right\}  = 
        \\
        &= \beta^{|R|\over s} U^{(s,1)}_R \left( p_k =  N \delta_{k|3}  + \delta_{k-1|3} + \delta_{k-2|3}  \right) 
\end{split}
\end{equation}
Keep $a=0$ and let $N=2 \mod 3$, then:
\begin{equation}
    p_k = \sum_{i=1}^{N} \omega_e^{(i-1)k} = \left\{ \begin{split}
         &N \, , \, k=0 \mod 3
         \\
         & e^{\pi i \over 3} \, , \, k=1 \mod 3
         \\
          & e^{-{\pi i \over 3}} \, , \, k=2 \mod 3
    \end{split}
    \right. =    N \delta_{k|3}  + e^{\pi i \over 3}\delta_{k-1|3} +  e^{-{\pi i \over 3}}\delta_{k-2|3} 
\end{equation}
\end{enumerate}
In the generic case the corresponding locus in terms of $p_k$ variables will be given by multiple root of unity sums, while it looks much simpler in the $x_i$ variables. Formulas for $[[\beta N - \beta i+j]]_{s,a}$ can be obtained similarly. One way to do it is to switch to $p_k$ variables, where one can formally substitute $N$ by $N-1+\frac{1}{\beta}$, i.e.:
\begin{equation}
    \left( \prod_{(i,j) \in R}[[\beta N - \beta i+j)]]_{s,0} \right) U^{(s,1)}_R\left\{ \delta_{k,s} \right\}  = \beta^{|R|\over s} U^{(s,1)}_R \left( p_k = \delta_{k|s} \left(N-1+\frac{1}{\beta}\right) \right) 
\end{equation}
One can get the corresponding formulas for Schur functions simply by taking the $\beta \rightarrow 1$ limit. Therefore, in this section we have demonstrated that the content products of the non-Gaussian model can be represented as Uglov/Schur polynomials are special points restoring a stronger version of superintegrability in this model. One important point to make here is that this statement is actually also in the spirit of the next section. The fact that the simplest $\beta$-deformation of the non-Gaussian content product requires precisely Uglov polynomials is a strong argument in favor of the validity of the model.

\section{In support of the conjecture}\label{sec:Support}
In this section, we provide various arguments for the validity of our conjecture and the relevance of the root of unity limit for non-Gaussian matrix models. In particular, in sec. \ref{sec:Motivation} we explain the main reasoning that brought us to consider this conjecture.
\subsection{Explicit evaluation of the integrals}
The most straightforward way to try to test or look for superintegrability is to compute expectation values directly at fixed $N$ and $\beta$ and try to come up with the proper symmetric functions. Consider the simplest, not trivial example is the one with parameters $r = 2, s = 2$. First, notice that averages in odd degree vanish. Therefore, it makes sense to compute only correlators with even degree. A simple basis to do the calculations initially is the power sums basis. Therefore at level two we get
\begin{equation}
    \ev{ p_{2} = \sum_{i=1}^N x_i^2}^{(2,1)}_{2,1}  = (1-\beta)N +\beta N^2
\end{equation}
Since it is quadratic in $N$ we only need to compute it at three points. Next we compute $\ev{p_1^2}$ to obtain
\begin{equation}
    \ev{ p_{1}^2= \left(\sum_{i=1}^N x_i \right)^2 }^{(2,1)}_{2,1}   =
        (N-1)\beta+1 \, , \text{ for odd } N \quad \ev{ p_{1}^2} = N\beta \, , \text{ for even } N
\end{equation}
Here we start seeing the dependence on the parity of $N$. Even though in each case the $N$ dependence is linear, we need at least four points to interpolate this result. The key problem is that by increasing $N$ the complexity of the calculation grows. This is mostly because of the measure part which in the Uglov case involves $2N(N-1)$ brackets that have to be expanded. And for bigger $\beta$ each bracket also produces many terms.

As we can see the result strongly depends on parity of $N$, at the next level we already know to take the parity into account. And compute only for odd $N$. This actually makes us go even to higher values of $N$ as now we need to get up to $4$ points. Anyhow, we can compute \footnote{In fact for some of the correlators we used the help of Virasoro constraints, which we explain later, however, using so more computational power, one could power through the level $4$ case}:
\begin{equation}
\begin{split}
    &\ev{p_4}^{(2,1)}_{2,1}  = \left(-\beta +\beta  N+1\right) \left(2 \beta  N^2-2 \beta  N+2 N+1\right) 
    \\
    &    \ev{p_2^2}^{(2,1)}_{2,1} =  N \left(-\beta +\beta  N+1\right) \left(\beta  N^2-\beta  N+N+2\right)
   \\
   & \ev{p_2 p_1^2}^{(2,1)}_{2,1}  =\left(\beta  \left(N-1\right)+1\right) \left(\beta  \left(N-1\right) N+N+2\right)
   \\
   & \ev{p_1^4 }^{(2,1)}_{2,1}  =3 \left(-\beta +\beta  N+1\right)^2 
    \\
   & \ev{p_3 p_1 }^{(2,1)}_{2,1}  = 3 \left(1 - \beta + \beta N\right)^2 
\end{split}
\end{equation}
Using these results we can try to guess the correct linear combinations that should give factorized answers. Level 4 is already quartic in $N$, which makes it a non-trivial task. One can immediately see that taking the standard Jack polynomials doesn't work. For example, taking $R=[4]$ one gets:
    \begin{equation}
\begin{split}
    \ev{U^{(1,1)}_{[4]}}^{(2,1)}_{2,1}  &= \frac{3 (-\beta +\beta  N+1)}{(\beta +1) (\beta +2) (\beta +3)} \times  \\
    & \hspace{1.5cm}\times  \left(-\beta ^4+\beta ^3-4 \beta ^2+8 \beta +\beta ^2 N^3+2 \beta ^3
   N^2-\beta ^2 N^2+ \right.
   \\
   & \hspace{3cm} \left. +5 \beta  N^2+\beta ^4 N-2 \beta ^3 N+10 \beta ^2 N-2 \beta  N+4
   N+2\right)
\end{split}
\end{equation}
This expression quite explicitly demonstrates that superintegrability is a rather non-trivial statement: choosing the wrong basis leads to completely intractable formulas. In the case at hand it's possible to make the correct guess and deduce the level $4$ Uglov polynomials:
\begin{equation}
    \begin{split}
        & \ev{U^{(2,1)}_{[4]}}_1 = \frac{(\beta N+1) (\beta N+3) (\beta N- \beta+1) (\beta N- \beta+3)}{(\beta+1) (\beta+3)} \\
        & \ev{U^{(2,1)}_{[3, 1]}}_1 = -\frac{\beta (N-1) (\beta N+1) (\beta N-\beta+1) (\beta N-\beta+3)}{2 (\beta+1)^2}\\
        & \ev{U^{(2,1)}_{[2, 2]}}_1 = \frac{(N-1) (\beta N+1) (\beta N-2 \beta+2) (\beta N-\beta+1)}{2 (\beta+1)}\\
        & \ev{U^{(2,1)}_{[2, 1, 1]}}_1 = -\frac{(N-1) (\beta N+1) (\beta N-3 \beta+1) (\beta N-\beta+1)}{2 (3 \beta+1)}\\
        & \ev{U^{(2,1)}_{[1, 1, 1, 1]}}_1=\frac{1}{8} (N-3) (N-1) (\beta N-3 \beta+1) (\beta N-\beta+1)
    \end{split}
\end{equation}
However, further, for bigger partitions and for higher $s$ it becomes way too complicated. Calculating at bigger $N$ is computationally challenging and blindly guessing the factorized formulas is hard. However, as we explain later we actually know from symmetry grounds that Uglov polynomials should indeed be the correct basis. Now we only need to check this conjecture. This is simpler. One thing we can do is to fix a low enough $N$, say $N=3$ and calculate the averages of Uglov polynomials at varying $\beta$. In that manner we should still obtain formulas which are fully factorized in $\beta$. Then it is also simple to interpolate the $N$ dependence, since we conjecture, that in each factor the $N$ dependence is linear as well. In that way we are only bound be computing Uglov polynomials, and by the length of the partition, since for given $N$ only polynomials with $l(R) \leq N$ are non-vanishing.

\subsection{Superintegrability at $v=0$.}

It appears that at $v=0$ the model can be reduced to the standard $\beta$-deformation at least when the expectation values are non-zero. Therefore, in the non-vanishing sector of the $v=0$ model superintegrability can be proved. As stated above, non-zero expectation values correspond to $v=0$ Uglov polynomials for partitions with all parts divisible by $s$. Consider such a case:
\begin{equation}
    \ev{U^{(s,0)}_R(x_i)}_a =  \int\limits_{C^{\otimes N}_{s,a}} \prod_{i=1}^N dx_i \prod_{i<j}(x_i^s-x_j^s)^{\frac{2 \beta}{s}}\exp\left( - \sum\limits_{i=1}^N \dfrac{x_i^s}{s} \right) U^{(s,0)}_R(x_i)\, . 
\end{equation}
This requires $R=s \cdot\Tilde{R}$, where the notation means that $R_i=s \cdot \tilde{R}_i$, such that $\tilde{R}$ is an integer partition. Now we make a substitution:
\begin{equation}
    x_i \rightarrow ( x_i)^{\frac{1}{s}} \, \quad \text{and} \quad \beta= s \cdot \tilde{\beta}
\end{equation}
Therefore we obtain, using \eqref{eq:v0UglovtoJack} for the Uglov polynomials we obtain:
\begin{equation}
     \ev{U^{(s,0)}_R(x_i)}_a =\delta_{a,1} \cdot  \int\limits_{0}^{\infty} \prod_{i=1}^N dx_i  x_i^{\frac{1-s}{s} } \prod_{i<j}(x_i-x_j)^{2 \tilde{\beta}}\exp\left( - \sum\limits_{i=1}^N \dfrac{x_i}{s} \right) J_{\tilde{R}}(\tilde{\beta}|x_i) \, .
\end{equation}
After this change of variables all the rays are ''glued'' together:
\begin{equation}
    B_{s,j} \quad \rightarrow \quad (0,\infty)
\end{equation}
for any $j$, hence the total contour becomes proportional to the $(0,\infty)$ half-line:
\begin{equation}
    C_{s,a} \quad  \rightarrow  \quad \left( \sum_{j=0}^{s-1}\omega_s^{-j (a-1)} \right) \cdot 
    \left( 0, \infty \right) =s\, \delta_{a,1}   \cdot 
    \left( 0, \infty \right)
\end{equation}
Here an additional power of $\omega^{-b}$ appears because of the change of variables in the integration. The resulting model is noting but the $\beta$-deformed complex $N \times N$ matrix model or the Wishart-Laguerre $\beta$-ensemble. It has an additional factor of $x_i^{\frac{1-s}{s} }$ , which in these models is interpreted as a determinant insertion. The model actually enjoys superintegrability at any value of the exponent of this insertion \cite{Cassia:2020uxy}:
\begin{equation} \int\limits_{0}^{\infty} \prod_{i=1}^N dx_i\,  x_i^{\nu} \prod_{i<j}(x_i-x_j)^{2\beta}\exp\left( - a_1 \sum\limits_{i=1}^N x_i \right) J_{R}(x_i)  = \dfrac{J_R \left\{\beta^{-1}(\nu+\beta(N-1)+1) \right\}}{J_R \left\{ \beta^{-1} a_1 \delta_{k,1}\right\}}  J_R \left\{ N \right\}
\end{equation}
The integral is convergent for $Re(\nu) > -1$, which in our case is satisfied. Therefore, substituting $\nu=\frac{1-s}{s}\,,a_1 =\frac{1}{s}$ and $\beta \rightarrow \tilde{\beta}, R \rightarrow \tilde{R}$ we can present :
\begin{equation}
    \begin{split}
        J_{\tilde{R}} \left\{\beta^{-1}(\nu+\beta(N-1)+1) \right\} &= \prod_{(i,j) \in \tilde{R} } \left( \nu+ \tilde{\beta} (N-1)+1 - \tilde{\beta} (i-1) + (j-1)  \right) \cdot J_{\tilde{R}} \left\{ \tilde{\beta}^{-1} \delta_{k,1}\right\}
    \\
    J_{\tilde{R}} \left\{N \right\} &= \prod_{(i,j) \in \tilde{R} } \left( \tilde{\beta} N - \tilde{\beta} (i-1) + (j-1)  \right) \cdot J_{\tilde{R}} \left\{ \tilde{\beta}^{-1} \delta_{k,1}\right\}
    \end{split}
\end{equation}
These products, taken by boxes of $\tilde{R}$, can be rewritten in terms of the diagram $R$:
\
\begin{equation}
    \begin{split}
        &\prod_{(i,j) \in \tilde{R} } \left( \tilde{\beta} N - \tilde{\beta} (i-1) + (j-1)  \right)  = s^{-|\tilde{R} |}\prod_{(i,j) \in \tilde{R} } \left( \beta N - \beta (i-1) + s (j-1)  \right) = 
        \\
        &\hspace{8.5cm}=s^{-|\tilde{R} |}\prod_{ \substack{(i,j) \in R \\ j = 1 \mod s } } \left( \beta N - \beta (i-1) +  (j-1)  \right)
        \\
        &\prod_{(i,j) \in \tilde{R} } \left( \nu+ \tilde{\beta} (N-1)+1 - \tilde{\beta} (i-1) + (j-1)  \right) = 
        \\
        &\hspace{4cm}=s^{-|\tilde{R} |} \prod_{(i,j) \in \tilde{R} } \left( 1 + \beta (N-1) - \beta  (i-1) + (j-1) s  \right) =\\
        &\hspace{8cm}= s^{-|\tilde{R} |} \prod_{ \substack{(i,j) \in R \\ j = 1 \mod s }  } \left(  \beta (N) - \beta  i + j  \right)
    \end{split}
\end{equation}
We see that the polynomial part of these expressions completely coincides with what we should have obtained according to the second part of \eqref{eq:result2}. We also need only to check if the product of the $N$-independent coefficients match:
\begin{equation}
    \frac{s^{-2|{\tilde{R}}|}J_{\tilde{R}} \left\{ \tilde{\beta}^{-1} \delta_{k,1}\right\}J_{\tilde{R}} \left\{ \tilde{\beta}^{-1} \delta_{k,1}\right\}}{J_{\tilde{R}} \left\{\beta^{-1} \delta_{k,1}\right\}} = J_{\tilde{R}}\left\{\beta^{-1} \delta_{k,1}\right\} = U^{(s,0)}_R\left\{\beta^{-1}\delta_{k, s}\right\} = \frac{U^{(s,0)}_R \{\delta_{k, s}\}}{\beta^{\frac{|R|}{s}}}
\end{equation}
Combining the results above, we obtain exactly the desired formula. 
\subsection{Virasoro constraints}
One of the powerful tools in the theory of matrix models are the Virasoro constraints \cite{Mironov:1990im,Dijkgraaf:1990rs,Cassia:2021dpd}. They can be presented as a set of relations between correlators or as differential equations for the times extended partition function. In both cases, their origin is the independence of the integral under the change of integration variables. Relation of Virasoro constraints to superintegrability was studied in \cite{Mironov:2021yvg}. However, that work was limited to a few simple cases. In the case at hand Virasoro constraints for generic potential were studied in \cite{Khlaif:2022mhv}. Already in this case it appears that, contrary to the standard story, only a certain subset of constrains can be constructed. On the other hand, it is well known, that when one specifies the potentials, additional features might appear. This specifically happens in the cases of non-Gaussian monomial (and polynomial) potentials. In that case even in the standard Hermitian models, the Virasoro constraints do not completely determine the partition function, but have ambiguities related to contour choices.
\\

Let us first briefly review the standard Hermitian case. To do this, introduce the partition function:
\begin{equation}\label{eq:timeZ}
    Z_N(T_k) = \int \prod_{i=1}^N dx_i  \Delta^2(x) \exp\left(\sum _{k=1}^{\infty} \dfrac{T_k}{k} \sum_{i=1}^N x_i^k \right)
\end{equation}
In this generic Hermitian matrix model we should specify a ''phase'', i.e. which of the times $T_k$ we treat as finite and which perturbatively as generating parameters. The monomial matrix models, for example, are obtained by a redefinition of times:
\begin{equation}\label{eq:shift}
    T_k  = \delta_{k,s} +p_k 
\end{equation}
Then the partition function \eqref{eq:timeZ} is a generating function of all correlation functions is in the corresponding models, where the time variables $p_k$ are treated as formal generating parameters. A relation to previous considerations is achieved by expanding this  generating function using the Cauchy identity \eqref{eq:Cauchy} as:
\begin{equation}\label{eq:CharExp}
    Z_N(p_k) = \sum_R \ev{S_R(x)} S_R(p)
\end{equation}
where the sum is over all partitions. We remind that the partition function treated as a function of $p_k$ satisfies Virasoro constraints. These are equations for the partition function that can be obtained as a consequence of the independence of the integral from infinitesimal variable shifts:
\begin{equation}
    x_i \rightarrow x_i +\epsilon x_i^{n+1}
\end{equation}
Which produces:
\begin{equation}
    L_n Z_N(p_k) = 0 \, , \quad n \geq -1
\end{equation}
Now, in the Gaussian model, these constraints completely fix the form of the partition function as a formal series, and therefore, by formula \eqref{eq:CharExp} determine the Schur expectation values. On the other hand, already in the cubic model this is not the case and the system of Virasoro constraints has a non-trivial kernel given by generic functions of $p_1$. This subject is covered in detail in \cite{Cassia:2021dpd}.
\\

Now let us move on to the Virasoro constraints in the Uglov model:
\begin{equation}\label{eq:timeZ}
    Z^{(\text{Uglov})}_N(T_k) = \int \prod_{i=1}^N dx_i  \Delta_{\text{Uglov}}^{(s,1)}(x) \exp\left(\sum _{k=1}^{\infty} \dfrac{T_k}{k} \sum_{i=1}^N x_i^k \right)
\end{equation}
The derivation for a generic potential, just as in \eqref{eq:timeZ}, was done in \cite{Khlaif:2022mhv}. The key takeaway for us is that due to the structure of the Uglov determinant, only a subset of equations can be obtained. In particular, for the constraint generated by  $x_i \to x_i + \epsilon x_i^{n+1}$ the variation of the Uglov determinant contains:
\begin{equation}
    \begin{split}
     (x_i^r&-x_j^r)^{2k} \longrightarrow (x_i^r+\epsilon r x_i^{n+r}-x_j^r-\epsilon r x_i^{n+r})^{2k} \approx \\ 
     &\approx (x_i^r-x_j^r)^{2k}+2k\epsilon r(x_i^r-x_j^r)^{2k-1}(x_i^{n+r} - x_j^{n+r})
    = (x_i^r-x_j^r)^{2k}\left( 1+2kr\epsilon\frac{x_i^{n+r} - x_j^{n+r}}{x_i^r - x_j^r} \right)
\end{split}
\end{equation}
where $k = \frac{(\beta-1)}{r}$. This can be reproduced by a differential operator in times variables  only if $n$ is divisible by $r$. Hence, we obtain:
\begin{equation}
    L^{\text{Uglov}}_{n r} Z^{(\text{Uglov})}_N(T_k) = 0 \, ,\quad n \geq 0
\end{equation}
For generic couplings one obtains:
\begin{equation}\label{eq:VirasoroUglov}
    \begin{split}
            & L^{\text{Uglov}}_{n r} =
            \sum_{i=1}^{\infty}(i+nr)T_i\frac{\partial}{\partial T_{i+nr}} + \sum_{i=1}^{nr-1} i(nr-i)\frac{\partial^2}{\partial T_i \partial T_{nr-i}} + +(2N + kr(2N - n - 1))nr\frac{\partial}{\partial T_{nr}}\\& 
    +kr\sum_{i=1}^{n-1} ir(nr-ir)\frac{\partial^2}{\partial T_{ir} \partial T_{nr-ir}} 
    + \delta_{n, 0}(N^2+krN(N-1))           
    \end{split}
\end{equation}
The Virasoro constraints for the monomial Uglov model are obtained by a redefinition of variables similar to \eqref{eq:shift}. However, now, to accommodate for the expansion in terms of Uglov polynomials, it also makes sense to normalize the times in such a way that we obtain the correct Cauchy formula \eqref{eq:CauchyUglov}. That is, we set in the Virasoro constrains \eqref{eq:VirasoroUglov}:
\begin{equation}
    T_k  =  \delta_{k,s} + \beta^{\delta_{k|s}} p_k 
\end{equation}
Therefore we set $r=s$ and rewrite the Virasoro constraints in terms of the  $p_i$ variables:
\begin{equation}\label{eq:VirasoroUglovFull}
    \begin{split}
            & L^{\text{Uglov}}_{n s} = -\frac{(ns+s)}{\beta}\frac{\partial}{\partial p_{ns+s}}+
            \sum_{i=1}^{\infty}(i+ns)p_i\frac{\partial}{\partial p_{i+ns}} + \sum_{i=1}^{ns-1} \beta^{-2\delta_{i|s}}i(ns-i)\frac{\partial^2}{\partial p_i \partial p_{ns-i}} +
            \\& + (2N + (\beta-1)(2N - n - 1))\frac{ns}{\beta}\frac{\partial}{\partial p_{ns}}
    +\beta^{-2}(\beta-1)\sum_{i=1}^{n-1} ir(ns-is)\frac{\partial^2}{\partial p_{is} \partial p_{ns-is}} 
    + 
    \\& +\delta_{n, 0}(N^2+(\beta-1)N(N-1))           
    \end{split}
\end{equation}
These equations allow for a very nontrivial check of our conjecture. Namely, we check:
\begin{equation}
   L^{\text{Uglov}}_{n s} \cdot \sum_R \dfrac{\ev{U^{(s,1)}_R}_a U_R^{(s,1)}} {||U^{(s,1)}_R||^2 }  = 0 \,  , \quad  n \geq 0
\end{equation}
Here $N$ and $\beta$ can be treated simply as parameters, and we are not forced to interpolate between discrete values. As a result, we are bounded only by complexity of calculating Uglov polynomial, and hence Macdonald polynomials, for bigger representations.  We checked our conjecture from $r=2$ to $r=5$ at least for $|R| \leq 10$. Surely one can go even to higher orders.
\\

\subsection{$W$-representation}\label{sec:Motivation}
Another important piece of evidence for the conjecture, and the main source of intuition for us, is the consideration of relevant $W$-operators. In this section, we will briefly remind the main construction of $W$-operators for matrix models and then see where it leads when applied to the monomial matrix models. The main idea behind $W$-operators is that they are a sort of hidden symmetry of the matrix model, that is supposedly responsible for superintegrability. This point of view is explored in \cite{Mironov:2020pcd}, where it was also explained how to construct $W$-operators for specific models. The starting point of this subject is the observation \cite{Morozov:2009xk} that times dependent partition functions of some matrix models can be represented in the form:
\begin{equation}
    Z(p_k) = \exp\left( \hat{W}\left[ p_k ,\dfrac{\partial}{\partial p_k}\right] \right) \cdot 1
\end{equation}
The operator $\hat{W}$ acts on parameters $p_k$ and depends on the model: on the potential, measure, and type of matrices integrated. In some cases it is either more appropriate or necessary that the exponent acts not on an identity, but on some other functions like in the WLZZ models \cite{Wang:2022fxr,Mironov:2023pnd} where it acts on an exponent with additional parameters.
Recently some progress was made in understanding the role of $W$-operators in superintegrability, and it was revealed that one should study their action on the relevant symmetric functions. In this regard one should view these $W$-operators as parts of some large algebra that has distinguished representations in terms of partitions. Such algebras, now studied under the name of BPS-algebras, include the $\mathcal{W}_{1+\infty}$ algebra, affine Yangians $Y(\hat{\mathfrak{gl}}_n)$ etc. \cite{Galakhov:2023gjs,Harvey:1996gc,Li:2020rij,Tsymbaliuk:2014fvq,Prochazka:2015deb,Schiffmann:2012tbu,Litvinov:2020zeq}. In regard to matrix models only some of these algebras have yet appeared. As a result of this section we will observe the relevance of supposedly the $Y(\hat{\mathfrak{gl}}_n)$ Yangian.
\\

It should be noted that this representation can in some  cases be derived from Virasoro constraints (or higher $W$-constraints). In these situations the aforementioned action on partitions can be used to prove superintegrability, \emph{à la} \cite{Mishnyakov:2022bkg}. In other cases it is not possible, or rather more peculiar, for instance, when Virasoro constraints do not completely fix the partition functions. The monomial and Uglov models are exactly of this sort. Nevertheless, classic techniques combined with recent progress on the matter allow reversing engineer the $W$-representation if superintegrability is already known, which then opens up the possibility to use methods coming from representation theory of relevant algebras. This is exactly the approach we are going to take here. 
\\

Returning back to monomial matrix models, the starting point is the undeformed case. Here superintegrability holds for Schur functions. Hence the relevant algebra of operators is the $\mathcal{W}_{1+\infty}$ algebra. In this case the following is known. Suppose we have the character expansion of the following form:
\definecolor{royalblue(traditional)}{rgb}{0.0, 0.14, 0.4}
\definecolor{spirodiscoball}{rgb}{0.06, 0.75, 0.99}
\definecolor{tangerine}{rgb}{0.95, 0.52, 0.0}
\begin{equation}
    Z(g(x),n \, | p_k) = \sum_R \prod_{(i,j) \in R} {\color{spirodiscoball}g(j-i)} S_R(\delta_{k,{\color{tangerine}n}}) S_R(p)
\end{equation}
Let us remind that these appear in matrix models, for example, as in \eqref{eq:CharExp}, when using the standard Cauchy formula.
Then one has a corresponding operator in the single free boson representation of the $\mathcal{W}_{1+\infty}$-algebra, such that:
\begin{equation}\label{eq:Wgeneric}
\begin{split}
     &Z(g(x),n \, | p_k) = \exp\Big( \hat{W}_{\color{tangerine}n}[{\color{spirodiscoball}g}] \Big) \cdot 1 
     \\
     &\hat{W}_{\color{tangerine}n}[{\color{spirodiscoball}g}] S_R(p)  = \sum\limits_{\substack{Q: \vspace{2pt} \\  |Q|=|R|+{\color{tangerine}n}}} \left(\prod_{(i,j) \in Q/R} {\color{spirodiscoball}g(i-j)} \right) \ev{ p_{\color{tangerine}n} S_R \big| S_Q} S_Q(p)
\end{split}
\end{equation}
where the coefficients $\ev{ p_n S_R \big| S_Q}$ can be computed via the scalar product on Schur polynomial and represent the action of the simplest examples $W$-operators - multiplication by $p_n$:
\begin{equation}
    p_n S_R =\sum\limits_{\substack{Q: \\ |Q|=|R|+n}} \ev{ p_n S_R \big| S_Q} S_Q
\end{equation}
We have highlighted in color the components that are in correspondence in the character expansion and in the action of the $W$-operator on Schur functions. One can describe the element \eqref{eq:Wgeneric} in the $\mathcal{W}_{1+\infty}$ algebra. For details and definition we refer to \cite{Mironov:2020pcd}. The key element is the operator:
\begin{equation}\label{eq:W0generic}
    \hat{W}_0[g] S_R =\left( \sum_{(i,j) \in R}  g(j-i) \right) S_R
\end{equation}
It should be thought as the building block for the whole set of operators \eqref{eq:Wgeneric}. After  \eqref{eq:W0generic} is provided one immediately constructs the whole family, by utilising iterated commutators, a construction explained in \cite{Mironov:2020pcd}.
\\

Let us treat the specific case we have at hand. The content function is given by the double-brackets:
\begin{equation}\label{eq:gMonomial}
    g(x)= [[N+x]]_{s,0}[[N+x]]_{s,a}
\end{equation}
In that form, this function is only defined for integer $x$, however, to find the $W$-operator we need to analytically continue it for generic $x$. This can be achieved by utilizing the fact that for integer $x$ one has:
\begin{equation}
    \sum_{k=0}^{s-1} \exp\left( \dfrac{2 \pi i k}{s} (x-a) \right)   = \left\{ 
    \begin{split}
        &1 \, , \, x\!\! \mod s = a
        \\
        &0 \, , \, x\!\! \mod s \neq a
    \end{split}
    \right.
\end{equation}
which can be represented as the value of the expression:
\begin{equation}
    \frac{1-q^{s(x-a)}}{1-q^{x-a}}=\sum_{k=0}^{s-1} q^{k(x-a)} 
\end{equation}
at $q = \omega_s$. At this point we can already observe the root of unity limit emerging, which in fact this is the key observation of this section. Now we can go on to explicitly represent the function \eqref{eq:gMonomial} as well and then build the $W$-operator. We will not go on to provide a full expression, as we don't yet have a concise way to present is. However, it is clear, that the building block of this $W$-operator will be an operator acting on Schur functions as:
\begin{equation}
    \hat{W}_{0}\left[q^{kx} \right] S_R = \left(\sum_{(i,j) \in R} q^{k(j-i)} \right) S_R
\end{equation}
Its one body representation is
\begin{equation}
    \hat{W}^{\text{one-body}}_0\left[q^{kx} \right]  = \dfrac{1-q^{kD}}{1-q^k}
\end{equation}
and hence we can explicitly write its second quantized bosonic representation:
\begin{equation}
    \hat{W}_{0} \left[q^{kx} \right]= \oint \dfrac{dz}{z} : e^{-\phi(z)} \dfrac{1-q^{kD}}{1-q^k}  e^{\phi(z)} = \dfrac{1}{q^k-1} \oint \dfrac{dz}{z} \left(  :e^{\phi(q^k z)-\phi(z)}:  -1  \right)
\end{equation}
Which is nothing but the Macdonald operator (normalized) at $q=t$ \cite{Zenkevich:2014lca}. Therefore the $W$-operator for a specific monomial matrix model with be made out of iterated commutators of the Macdonald operator evaluated at:
\begin{equation}
    q=t=\omega_s
\end{equation}
At this locus the Macdonald operator always acts on Schur functions, since for any value of $q=t$ Macdonald polynomials reduce to Schur polynomials. Therefore, the symmetric function in these models are not sensitive to the specific point in the $(q,t)$-plane. However, we clearly see, that the operators are different and moreover, we could clearly expect that deformation around this point would depend on the specific value of $q$. Following this logic we are led to search for the $\beta$-deformation around the root of unity locus, which turns out to be the correct thing to do as demonstrated in the previous sections. Finally, let us note that this $\beta$-deformation of the root of unity point has recently been studied from the purely algebraic standpoint, at least, in the case of $r=2$. It was shown  that the resulting algebra is the affine Yangian $Y(\hat{\mathfrak{gl}}_2)$ and the respective analogs of cut-and-join operators were constructed \cite[eq.(97)-(100)]{Galakhov:2024mbz}. 
\\

We point out once again, that these considerations were not derived from Virasoro constraints, since in non-Gaussian examples it is not possible, or at least not known at the moment. Hence in this section we described the logic that was announced in fig \ref{fig:logic}. 

\subsection{Back to $(q,t)$. Limit of the $(q,t)$ deformed Gaussian and Wishart-Laguerre models}

Here we would to explore the possibility to derive not only the Uglov determinant, but the whole matrix model together with the potential from the $(q,t)$-deformed integrals. To do this, let us first recall that it is conjectured that the $(q,t)$-analogs of the Gaussian and  Complex matrix (RCM) models are also exactly solvable. To define the models we first recall some notations. 

In the $(q,t)$-deformed case integration over eigenvalues is substituted by Jackson integrals:
\begin{equation}
    \int_{-\xi}^{\xi} d_q x f(x) = \xi (1-q) \sum_{n=0}^{\infty} q^n \left( f(\xi q^k) + f(-\xi q^k) \right) 
\end{equation}
The $q$-Pochammer symbol is defined as:
\begin{equation}
    (z;q)_{\infty}= \prod_{i=0}^{\infty}(1-z q^i)
\end{equation}
Then, the $(q,t)$-deformed Gaussian matrix model is defined as \cite{Cassia:2020uxy,Morozov:2018eiq}:
\begin{equation}\label{eq:Gaussian-qt}
    \ev{f(x)}^{\text{Gauss}}_{(q,t)} = \int\limits_{-\xi}^{\xi} \prod_{i=1}^N d_q x_i^{\beta(N-1)} \prod_{i \neq j}\dfrac{\left( \dfrac{x_i}{x_j};q\right)_{\infty}}{\left(\dfrac{t x_i}{x_j};q \right)}_{\infty}  \prod_{i=1}^N\left( \dfrac{q^2 x^2_i}{\xi^2} ;q^2 \right)_{\infty}f(x)
\end{equation}
where the role of the Gaussian measure is played by the Pochammer symbol $(q^2 z^2/\xi^2  x_i ;q)_{\infty}$. The $(q,t)$ analog of the complex matrix model/Wishart-Laguerre (WL) model is also available and given by:
\begin{equation}\label{eq:RCM-qt}
    \ev{f(x)}^{\text{RCM/WL}}_{(q,t)} = \int\limits_{-\xi}^{\xi} \prod_{i=1}^N d_q x_i^{\beta(N-1)} \prod_{i \neq j}\dfrac{\left( \dfrac{x_i}{x_j};q\right)_{\infty}}{\left(\dfrac{t x_i}{x_j};q \right)}_{\infty}   \prod_{i=1}^N \left( \dfrac{q x_i}{\xi} ;q \right)_{\infty} f(x)
\end{equation}

It is conjectured \cite{Cassia:2020uxy,Mironov:2022fsr} that averages of Macdonald polynomials in these models are also exact, i.e. the models posses superintegrability:
\begin{equation}
\begin{split}
       \ev{\operatorname{Mac}_R(x)}^{\text{Gauss}}_{(q,t)}& =  \frac{\operatorname{Mac}_R\left\{\frac{\xi^k(1+(-1)^k)}{1-t^k}\right\} \cdot \operatorname{Mac}_R\left\{\frac{1-t^{k N}}{1-t^k}\right\}}{\operatorname{Mac}_R\left\{\frac{-\xi^{2k}}{1-t^k}\right\}}
   \\
   \ev{\operatorname{Mac}_R(x)}^{\text{RCM/WL}}_{(q,t)} & =  \frac{\operatorname{Mac}_R\left\{\frac{1-q^k t^{k(N-1)}}{1-t^k}\right\} \cdot \operatorname{Mac}_R\left\{\frac{1-t^{k N}}{1-t^k}\right\}}{\operatorname{Mac}_R\left\{\frac{\xi^k}{1-t^k}\right\}}  
\end{split}
\end{equation}
For our purposes it is instructive to recall, for example, in the WL, how the undeformed model is obtained as a limit. We take the $q \rightarrow 1$ limit, while also properly scaling $\xi$. To recover the proper scaling, notice that:
\begin{equation}
    \prod_{i=1}^N \left( \dfrac{q x_i}{\xi} ;q \right)_{\infty} = \exp\left( - \sum_{k=1}^{\infty } \dfrac{1}{ k \xi^k (q^{-k}-1)} \sum\limits_{i=1}^N x_i^k    \right)
\end{equation}
Hence we rescale $\xi=\dfrac{q}{1-q}$, then: 
\begin{equation}
    \xi \rightarrow \infty  \, , \quad  \dfrac{(1-q)^k}{ k (1-q^{k})}  \, \longrightarrow  \, \delta_{k,1}
\end{equation}
and
\begin{equation}
     \prod_{i=1}^N \left( \dfrac{q x_i}{\xi} ;q \right)_{\infty} \quad \longrightarrow \quad \exp\left( - \sum_{i=1}^N x_i \right) 
\end{equation}
Now, instead of this limit we could scale $\xi$ as:
\begin{equation}
    \xi = \dfrac{1}{s(q^{s}-1)^{1/s}}
\end{equation}
Then as $q \rightarrow \omega_s$ we have:
\begin{equation}
    \xi \sim \omega_s \cdot \infty \, , \quad -\frac{s \left(q^s-1\right)^{k/s}}{k \left(q^{-k}-1\right)} \, \longrightarrow  \, \delta_{k,s}
\end{equation}
And, hence:
\begin{equation}
     \prod_{i=1}^N \left( \dfrac{q x_i}{\xi} ;q \right)_{\infty} \quad \longrightarrow \quad \exp\left(  \sum_{i=1}^N x_i^s \right) 
\end{equation}
Hence we can consistently recover the monomial potential by correctly scaling $\xi$ in the root of unity limit. On the other hand we see, that the integration limits behave in a somewhat reasonable way, however, it's still unclear how to completely reconstruct the contours of \eqref{eq:result2}. 
\\

What this discussion suggests is that one can recover the Gaussian matrix model from the $(q,t)$-deformed WL ensemble \eqref{eq:RCM-qt} in the limit $q \rightarrow \omega_2$. On the other hand one take the standard $q\rightarrow 1$ limit of the Gaussian $(q,t)$ model\eqref{eq:Gaussian-qt} instead. By similar considerations, one would have:
\begin{equation}
    \left( \dfrac{q^2 x_i^2}{\xi^2} ;q^2 \right)_{\infty} \quad \longrightarrow \quad \exp\left(  \sum_{i=1}^N x_i^2 \right) 
\end{equation}
Therefore, we clearly can observe that we obtain two versions of the Gaussian matrix model. One with an Uglov determinant with $r=s=2$ and the other with $r=1 \,,\, s=2$. We have observed in sec. \ref{sec:Result} that both case are superintegrable, which is in agreement with the limits.

\section{Discussion and outlook}\label{sec:Outlook}

We would like to conclude by reiterating a few key features of the obtained result and speculate about some further directions and generalization. 
\\

One thing that we want to stress is that our result fills in an important gap in the list of superintegrable/exactly solvable matrix models. Moreover, from this perspective we obtained a rather interesting example. Suppose we were to look at the case at hand from the perspective of finding the right basis for a given measure and potential. Given the rather complicated form of the Uglov determinant, guessing an answer might seem improbable. Moreover, no other methods for solving the model work in this case, so we can't even calculate a number of averages by some analytic method and then try to combine those to obtain factorized formulas. Hence it is quite surprising that an answer does exist. On the other hand, we observe that for a given $s$ only a special potential produces nice formulas, further stressing that superintegrability is a peculiar property. 

It's intriguing that the relevant polynomials once again have lots of nice group theoretic properties, which allows to further stretch the initial statement about characters of \cite{Mironov:2017och,Mironov:2022fsr}. The result presented here heavily relies on these group theoretic properties, as we have demonstrated in sec \ref{sec:Motivation}.

Our considerations point to a few important questions and further research direction, among those are:
\begin{itemize}
    \item The extension of Uglov polynomials to other values of $v$. The instanton counting problem for ALE spaces and the consistency of the matrix model with this limit together with superintegrability  naturally leads to this question. Naively, however, this limit of Macdonald polynomials is ill-defined.  
    \item Proving superintegrability algebraically. As we have explained, our proposal can't be proven directly from Virasoro constraints, since there is no obvious way to obtain the $W$-operator from them. This is not merely a technical difficulty but a conceptual one. Virasoro constraints do not fix the integration contour, while superintegrability in the presented form seems to strongly depend on it. On the other hand, Virasoro constraints and $W$-operators are among the few tools for matrix models that survive $\beta$-deformation.
    \item A more technical, but still interesting question is the study of the $v=0$ Uglov polynomials. Even though those of them that are relevant from the matrix model perspective they appear to reduce to Jack polynomial, the others are non-trivial. Moreover, we did not demonstrate that their averages vanish.
    \item There are several technical extensions that are expected based on previous results. First is the case of non-normalized averages \emph{à la } \cite{Barseghyan:2022txq}. Clearly, we expect that a distinguished role will be played by Uglov polynomials for rectangular partitions, which interplays nicely with \cite{Bershtein:2022fkn}. Second is the extension towards bilinear correlators along the lines of \cite{Chan:2023lhx}.
    \item Finally, it would be interesting to complete the analysis of the limit of the $(q,t)$-deformed RCM/WL and Gaussian model. In particular it is important to understand how the star-like contours of \cite{Cordova:2016jlu} arise in this limit. Apart from that the Gaussian model has a feature of having two superintegrabilities and naturally one could ask if this holds in any other cases.
\end{itemize}

\section*{Acknowledgements}
We are grateful to A. Popolitov, A. Zhabin, N. Tselousov, D. Galakhov, A.Mironov and A.Morozov for useful discussions. Nordita is supported in part by NordForsk.
\appendix 
\section{Symmetric functions and notations}\label{sec:AppendixNotations}
We briefly introduce a few notations from the theory of symmetric functions and partitions that are used throughout the paper \cite{macdonald1998symmetric,fulton1997young}. 
\\

Integer partitions and Young diagrams are denoted by $R  = [R_1,R_2,R_3, \ldots]$. Superintegrability formulas are often given in terms of the so-called contents of the partitions. It is defined in terms of the coordinates of a box in the Young diagram $(i,j)$ with $i$ being the row number and $j$ the column number.  For example, the shaded box in the figure below has coordinates $(2,3)$:
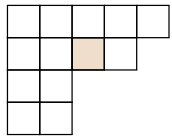
\vspace{0.3cm}
\begin{center}
    \ytableausetup{boxsize=1em,aligntableaux = center}  \quad \ydiagram{5,4,2,2}*[*(almond) ]{0,2+1}
\end{center}
\vspace{0.3cm}
The difference $j-i$ is called a content of a given box of a partition. Next, we illustrate the notations for special functions evaluated at special points on the example of Schur functions. Schur functions are characters of $GL(N)$ and can be computed in several ways. In particular, consider a generating function:
\begin{equation}
    \exp\left( \dfrac{z^k p_k}{k} \right) = \sum_k s_k(p) z^k
\end{equation}
The Schur polynomials are given by the determinant:
\begin{equation}
    S_R(p_k) = \det_{i,j} s_{R_i-i+j}(p_k)
\end{equation}
In this form, Schur functions are homogeneous functions of the variables $p_k$ of degree $R$, if $p_k$ is assigned a degree $k$. In symmetric functions notations $p_k$ are nothing but the powers sums:
\begin{equation}\label{eq:powersum}
    p_k = \sum_{i=1}^N x_i^k
\end{equation}
Making this substitution in Schur functions turns them into a symmetric functions of the $x_i$ variables. These can also be though as being eigenvalues of some matrix $X$, then one has:
\begin{equation}
    S_R(x_i)= S_R(p_k = \Tr X^k)
\end{equation}
In matrix models this is exactly how the Schur function appears in the integrand. On the r.h.s of superintegrability formulas we encounter Schur functions evaluated at special loci. Everywhere, except sec \eqref{sec:ContentProduct}, where it is spelled out explicitly, we mean special loci of the $p_k$ variables. For that we use a special notation, for example:
\begin{equation}\label{eq:not1}
    S_R\left\{ \delta_{k,s} \right\} := S_R(p_k = \delta_{k,s})
\end{equation}
which means we put all powers sums equal to zero except the $s$'th one, which is equal to one. We also use:
\begin{equation}\label{eq:not2}
     S_R\left\{ N \right\} := S_R(p_k = N )
\end{equation}
where all powers sums are equal to $N$. While the first point is natural only in the power sum basis, the second one be also expressed in the $x_i$, by putting all $x_i=1$. These special values of Schur functions can be described combinatorically as follows:
\begin{equation}
    \begin{split}
         &S_R\left\{ \delta_{k,s} \right\}   = \prod_{(i,j) \in R} \dfrac{1}{[[h_{i,j}]]_{s,0}} 
         \\
        &S_R\left\{N\right\} =  S_R\left\{ \delta_{k,1} \right\}  \prod_{(i,j)\in R} (N+j-i)
    \end{split}
\end{equation}
where $h_{i,j}$ are the corresponding hook length. i.e. the number of boxes right and below of the given one plus one, and the notation is as in \eqref{eq:doublebracket}.  With other symmetric functions that appear in the paper we use the same notation as in \eqref{eq:not1} and \eqref{eq:not2}, meaning that one expresses then in the power sum basis and evaluates at special values of the power sums.
\section{Uglov polynomials}\label{sec:AppendixUglov}
Here we list some more examples of Uglov polynomials for bigger partitions and various choices of $r$. 
$r=2, v=1$, $|R| \leq 4$:
\begin{equation}
    \begin{split}
        &U_{[1]}^{(2,1)} = p_1\\
        &U_{[1, 1]}^{(2,1)} = \frac{p_1^2}{2}-\frac{p_2}{2}\\
        &U_{[2]}^{(2,1)} = \frac{\beta p_2+p_1^2}{\beta+1} \\
        &U_{[1, 1, 1]}^{(2,1)} = \frac{p_1^3}{6}-\frac{p_2 p_1}{2}+\frac{p_3}{3} \\
        &U_{[1, 2]}^{(2,1)}=\frac{1}{3} \left(p_1^3-p_3\right)\\
        &U_{[3]}^{(2,1)}=\frac{3 \beta p_2 p_1+p_1^3+2 p_3}{3 \beta+3}\\
        &U_{[1, 1, 1, 1]}^{(2,1)} =\frac{p_1^4}{24}-\frac{1}{4} p_2 p_1^2+\frac{p_3 p_1}{3}+\frac{p_2^2}{8}-\frac{p_4}{4}\\
        &U_{[1, 1, 2]}^{(2,1)}=\frac{(2 \beta+1) p_1^4-3 (\beta+1) p_2 p_1^2-2 (\beta-1) p_3 p_1-3 \beta \left(p_2^2-2 p_4\right)}{18 \beta+6}\\
        &U_{[2, 2]}^{(2,1)}=\frac{3\beta p_2^2-3 (\beta-1) p_4+p_1^4-4 p_3 p_1}{6 (\beta+1)}\\
        &U_{[1, 3]}^{(2,1)}=\frac{(\beta+2) p_1^4+3 \beta (\beta+1) p_2 p_1^2+2 (\beta-1) p_3 p_1-3 \beta \left(\beta p_2^2+2 p_4\right)}{6 (\beta+1)^2}\\
        &U_{[4]}^{(2,1)}= \frac{3 \beta^2 p_2^2+6 \beta p_2 p_1^2+6 \beta p_4+p_1^4+8 p_3 p_1}{3 \beta^2+12\beta+9} 
    \end{split}
\end{equation}

$r=1$, $|R| = 4$ (Jack polynomials):
\begin{equation}
    \begin{split}
        &U_{[1, 1, 1, 1]}^1 = J_{[1, 1, 1, 1]} =\frac{p_1^4}{24}-\frac{1}{4} p_2 p_1^2+\frac{p_3 p_1}{3}+\frac{p_2^2}{8}-\frac{p_4}{4}\\
        &U_{[1, 1, 2]}^1=J_{[1, 1, 2]}=\frac{\beta p_1^4+(1-3 \beta) p_2 p_1^2+2 (\beta-1) p_3 p_1-p_2^2+2 p_4}{6 \beta+2}\\
        &U_{[2, 2]}^1=J_{[2, 2]}=\frac{\beta^2 p_1^4+\left(\beta^2+\beta+1\right) p_2^2-2 (\beta-1) \beta p_2 p_1^2-4 \beta p_3 p_1+(\beta-1) p_4}{2 (\beta+1) (2 \beta+1)}\\
        &U_{[1, 3]}^1=J_{[1, 3]}=-\frac{-\beta^2 p_1^4+(\beta-3) \beta p_2 p_1^2+2 (\beta-1) p_3 p_1+\beta p_2^2+2 p_4}{2 (\beta+1)^2}\\
        &U_{[4]}^1=J_{[4]}= \frac{\beta^3 p_1^4+6 \beta^2 p_2 p_1^2+8 \beta p_3 p_1+3 \beta p_2^2+6 p_4}{\beta^3+6 \beta^2+11 \beta+6}
    \end{split}
\end{equation}
$r=4, v=1$, $|R| = 4$:
\begin{equation}
    \begin{split}
        &U_{[1, 1, 1, 1]}^{(4,1)} =\frac{p_1^4}{24}-\frac{1}{4} p_2 p_1^2+\frac{p_3 p_1}{3}+\frac{p_2^2}{8}-\frac{p_4}{4}\\
        &U_{[1, 1, 2]}^{(4,1)}=\frac{(2 \beta+1) p_1^4-3 (\beta+1) p_2 p_1^2-2 (\beta-1) p_3 p_1-3 \beta \left(p_2^2-2 p_4\right)}{18 \beta+6}\\
        &U_{[2, 2]}^{(4,1)}=\frac{1}{12} \left(p_1^4-4 p_3 p_1+3 p_2^2\right)\\
        &U_{[1, 3]}^{(4,1)}=\frac{(\beta+2) p_1^4+3 (\beta+1) p_2 p_1^2+2 (\beta-1) p_3 p_1-3 \left(2 \beta p_4+p_2^2\right)}{12 (\beta+1)}\\
        &U_{[4]}^{(4,1)}= \frac{6 \beta p_4+p_1^4+6 p_2 p_1^2+8 p_3 p_1+3 p_2^2}{6 \beta+18}
    \end{split}
\end{equation}

\bibliographystyle{utphys}
\bibliography{Uglov}{}

\end{document}